\documentclass[hyper,12pt]{JHEP3}
\usepackage{epsfig}
\usepackage{amsmath}
\usepackage{amssymb}
\usepackage{epsf}
\usepackage{graphicx}
\usepackage{cite}
\usepackage{graphics}
\usepackage{epstopdf}

\newcommand\fverb{\setbox\pippobox=\hbox\bgroup\verb} \newcommand\fverbdo{\egroup\medskip\noindent% \fbox{\unhbox\pippobox}\
} \newcommand\fverbit{\egroup\item[\fbox{\unhbox\pippobox}]}
\newbox\pippobox

\newcommand{\nn}{\nonumber} 
\newcommand{\beq}{\begin{equation}}
\newcommand{\eeq}{\end{equation}} 
\newcommand{\beqa}{\begin{eqnarray}} 
\newcommand{\eeqa}{\end{eqnarray}}

\def\be{\begin{equation}}
 \def\ee{\end{equation}}
 \def\bea{\begin{eqnarray}}
 \def\eea{\end{eqnarray}}
 \def\bean{\begin{eqnarray*}}
 \def\eean{\end{eqnarray*}}

\newcommand{\ie}{{\it i.e.}}  \newcommand{\eg}{{\it e.g.}}

  \newcommand{\morder}[1]{{\cal O}\left(#1 \right)}
\newcommand{\eq}[1]{(\ref{#1})}

\newcommand{\half}{\frac{1}{2}}  
\newcommand{\lsim}{\lesssim} \newcommand{\gsim}{\gtrsim}

\def\COMMENT#1{}
\def\D{\displaystyle}
 
 \def\l{\left}
 \def\r{\right}

\def\bm#1{\mbox{\boldmath$#1$}}
 \def\slash{\hskip-0.7em/ \,}
 \def\esim{\,\mathrel{\rlap{\lower0.2em\hbox{$-$}}\raise0.15em\hbox{\footnotesize $\hskip0.04em\sim$}}\,}
 \def\gsim{\mathrel{\rlap{\lower0.2em\hbox{$\sim$}}\raise0.2em\hbox{$>$}}}
 \def\ksim{\mathrel{\rlap{\lower0.2em\hbox{$\sim$}}\raise0.2em\hbox{$<$}}}

\title{\center{Collisional Energy Loss of a Fast Muon \\ in a Hot QED Plasma.}}

\author{St\'ephane Peign\'e, Andr\'e Peshier\\ SUBATECH, UMR 6457, Universit\'e de Nantes \\ Ecole des
Mines de Nantes, IN2P3/CNRS. \\ 4 rue Alfred Kastler, 44307 Nantes
cedex 3, France \\ E-mail: \email{peigne@subatech.in2p3.fr,peshier@subatech.in2p3.fr}}

\abstract{We calculate the collisional energy loss of a muon of high energy $E$ in a hot QED plasma beyond logarithmic accuracy, 
\ie, we determine the constant terms of order $\morder{1}$ in $-dE/dx \propto \ln{E}+ \morder{1}$.
Considering first the $t$-channel contribution to $-dE/dx$, we show that the terms $\sim \morder{1}$ are sensitive 
to the full kinematic region for the momentum exchange $q$ in elastic scattering, including large values $q \sim \morder{E}$.
We thus redress a previous calculation by Braaten and Thoma, which assumed $q \ll E$ and could not find the correct constant 
(in the large $E$ limit). 
The relevance of 'very hard' momentum transfers then requires, {\it for consistency}, that $s$ and $u$-channel contributions 
from Compton scattering must be included, bringing a second modification to 
the Braaten-Thoma result. Most importantly, Compton scattering 
yields an additional large logarithm in $-dE/dx$. Our results might have implications in the QCD case of parton collisional 
energy loss in a quark gluon plasma.} 
 
\keywords{QED, plasma}

\begin{document}

\section{Introduction}

Jet quenching, as anticipated by Bjorken 25 years ago \cite{bj}, is a prominent 
signature of the intriguing state of matter created at the Relativistic Heavy
Ion Collider. The suppression of {\it light} hadron spectra at large transverse momentum $p_{\perp}$ \cite{phenix,star} can be explained -- at least qualitatively -- by attributing the hadron attenuation to the radiative energy loss of the parent parton (light quark or gluon) induced by its rescatterings in the hot or dense medium.
On the other hand, recent experimental data on heavy flavour quenching 
\cite{Adler:2005xv,Bielcik:2005wu}, measured indirectly via the $p_{\perp}$-spectra of electrons from $D$ and $B$ meson decays, suggest that the radiative energy loss of {\it heavy} quarks might be insufficient to explain the observed attenuation\footnote{It is stressed in Ref.~\cite{Armesto:2005mz} that this statement might be somewhat premature, since the theoretical calculation of heavy quark production suffers from large uncertainties already in proton-proton collisions, and also because the contributions to the electron spectra from charm and beauty
are not separated experimentally.}. 
This renewed the interest in the {\it collisional} part $-\Delta E_{coll}$ 
of the parton energy loss \cite{Wicks:2005gt}, which in the case of a heavy quark might not be negligible -- contrary to what has often been assumed. 
A basic quantity required to estimate collisional quenching is the rate of energy loss per unit distance, $-dE/dx$, of a parton produced in the remote past and travelling in a large size medium, as studied in Refs.~\cite{bj,TG,BTqcd}. 
For heavy ion collisions, where a parton initially produced in a hard subprocess crosses a medium of finite size $L$, we expect deviations from the linear law $-\Delta E_{coll}(L) = (-dE/dx) \cdot L$ \cite{PGG,Djordjevic:2006tw,Wang:2006qr,Gossiaux:2006yr,Gossiaux:2007gd}. However, the knowledge of $-dE/dx$ is a prerequisite before attempting any evaluation of $-\Delta E_{coll}$.

So far, the most detailed calculation of $-dE/dx$ for a heavy quark in the quark gluon plasma is done by Braaten and Thoma \cite{BTqcd}, and is based on their previous evaluation of muon collisional energy loss in QED \cite{BT}.
Here we will reconsider the latter calculation, which appears to suffer from an incorrect assumption on the magnitude of the momentum exchange in elastic scattering. 
As in Ref.~\cite{BT} (referred to as BT in the following) we study the propagation of a muon of mass $M$ and momentum $P = (E, \bm p)$ in an $e^\pm\gamma$ plasma at a temperature $T \ll M$, but large enough to neglect the electron mass.
The muon can be considered as a test particle, losing (or gaining) energy due to elastic ($2\rightarrow2$) scattering off thermal particles of momentum $K$. The latter can be electrons and positrons ($t$-channel scattering, see Fig.~\ref{fig:SE}a), or thermal photons (Compton scattering, see Fig.~\ref{fig:compton}). 
The Mandelstam invariants of the elastic processes are defined as
\beq
  \label{mandelstam}
  s = (P+K)^2 \, , \ \ t = (P-P')^2 \, , \ \ u =(P-K')^2 \, \ .
\eeq
In the high energy limit\footnote{When $E$ becomes large, $s = M^2 +2 PK \simeq 2 PK \sim \morder{ET}$, since $K \lsim \morder{T}$ is constrained by a thermal distribution. The high energy limit $s \gg M^2$ is thus equivalent to $E \gg M^2/T$. For convenience it will often be referred to as the $v \equiv p/E \to 1$ limit.}, the BT calculation {\it incorrectly} assumes that the (spacelike) momentum exchange $P-P' \equiv Q = (\omega,\bm q)$ is always small compared to the incoming muon energy, namely $|\omega| \leq q \equiv |\bm q| \ll E$. 
%SP 
We mention that this problem has also been noted recently in the QCD context of Refs.~\cite{Zakharov,Qin}\footnote{Those
references indeed mention the need for a careful treatment of the kinematics, including the region of large 
transfers $q \sim \morder{E}$. However, those studies seem to focus on $t$-channel scattering. In the present paper
we also stress (in the case of QED), that including Compton scattering is required for consistency.}.
%SP

As a first consequence of the BT assumption $q \ll E$, the Compton contribution $-dE_{\gamma}/dx$ was neglected in Ref.~\cite{BT}.
Second, in the $v \to 1$ limit the result for the energy loss $-dE_e/dx$ due to scattering off electrons (and positrons) found in \cite{BT},
\beq
\left. -\frac{dE_e}{dx} \right|_{{\rm BT}}^{v\to 1} = \frac{e^4 T^2}{48 \pi}
\left[ \ln{\frac{2 E}{e^2 T}}  + 2.031  \right] ,
\label{BTfull}
\eeq
has an incorrect constant next to the leading logarithm.
For a heuristic argument showing the importance of large momentum exchange $q$, we note that the logarithmic energy dependence in \eq{BTfull} arises from an integral $\int_T^{q_{\rm max}} dq /q$, with $q_{\rm max} \simeq E$ at high energy.
While the region $q \sim q_{\rm max}$ does not contribute to the leading logarithm, it {\it does} contribute to the constant next to it (\eg, the interval $[q_{\rm max}/2, q_{\rm max}]$ yields $\ln{2}$). 
From this simple observation we infer that the approximation $q \ll E$ used in the BT calculation is legitimate only at logarithmic accuracy, but not to calculate the constant term in \eq{BTfull}. The evaluation of this constant requires an accurate treatment of the very hard region $q \sim q_{\rm max}$. For the $t$-channel contribution we will derive the following analytic result correcting \eq{BTfull},
\beqa
\left. -\frac{dE_e}{dx} \right|^{v\to 1} &=& \frac{e^4 T^2}{48 \pi}
\left[ \ln{\frac{2 E}{e^2T}}  + \underbrace{\ln{24}  - \gamma + \frac{\zeta'(2)}{\zeta(2)}} - \frac{3}{4} \right]
\simeq \frac{e^4 T^2}{48 \pi} \left[ \ln{\frac{2 E}{e^2T}}  +  1.281 \right]  \label{PPfull} \, , \nn \\
&& \hskip 4cm 2.031
\eeqa
where $\gamma \simeq 0.577$ is Euler's constant.

We stressed that working {\it beyond} logarithmic accuracy requires
considering the kinematic region of very hard transfers $q \sim E$. Since $\bm k' = \bm k + \bm q$ and $k \lsim \morder{T}$, this corresponds to $k' \sim E$. Using
\beq
s = (P'+K')^2 = 2 P'\cdot K' + M^2 \mathop{\simeq}_{v\to 1} 2E' k' (1 -\cos{\theta_{p'k'}}) \, ,
\label{sestimate}
\eeq
we infer from $s \sim ET$ and $k' \sim E$ that the angle $\theta_{p'k'}$ between
$\bm p'$ and $\bm k'$ must be small, and that the constant next to the leading logarithm (partly) arises from the angular region
\beq
\theta_{p'k'} \sim \sqrt{T/E} \ll 1 \; .
\label{angles}
\eeq
Here is an essential point. When $E \to \infty$, the constant is sensitive to scatterings where $\bm p'$ and $\bm k'$ are collinear. But collinear outgoing particles should be associated within the same `jet' of particles, and such collinear
configurations should thus be removed from the definition of {\it observable} collisional energy loss.
Hence, we conclude that {\it when $E \to \infty$, the constant next to the leading logarithm in $-dE/dx$ is not an observable}, only the leading logarithmic term of \eq{PPfull} is meaningful. Strictly speaking, in the asymptotic limit $E \to \infty$ we can only state
\beq
 -\left. \frac{dE_e}{dx} \right|^{E\to \infty} \rightarrow \frac{e^4 T^2}{48 \pi}\,  \left[ \ln{\frac{E}{e^2 T}}  + \morder{1} \right] ,
\label{interpretation1}
\eeq
where the constant $\sim \morder{1}$ depends on the details of the jet definition.

For finite $E$, however, $\theta_{p'k'} \sim \sqrt{T/E}$ is a non-zero angle\footnote{For instance, for $E = 10 \,{\rm GeV}$
and $T= 500 \,{\rm MeV}$ we have $\sqrt{T/E} \simeq 0.22\,{\rm rad} \simeq 13^\circ$.}. Depending on the experimental angular resolution, processes transferring a large fraction of the incoming energy $E$ to the particle $K'$ (forming with $P'$ the angle $\theta_{p'k'} \neq 0$) might be counted as observable energy loss. In the present study we assume the angular resolution to be much better than $\sqrt{T/E}$.
Then it is meaningful to include the (correct) constant next to the leading logarithm, as done in \eq{PPfull}.

Comparing now \eq{BTfull} and \eq{PPfull}, it might seem that the difference in the constant term is only of minor importance.
However, the incorrect approximation of Ref.~\cite{BT} has also been used to calculate the collisional energy loss of a heavy quark in QCD \cite{BTqcd}, and the latter must thus also be corrected. 
Moreover, the relevance of the region $q \sim E$ in the $t$-channel contribution suggests that $s$ and $u$-channels (Compton scattering) are important, contrary to what is assumed in \cite{BT}.
Indeed, we find in the present case of QED (see section 3) that the Compton scattering contribution
precisely arises from the domain \eq{angles}\footnote{The presence of $s$ and $u$-channel contributions arising from the region $q \sim E$ will also affect the QCD results obtained in \cite{BTqcd}.}. 
Since quasi-collinear configurations \eq{angles} already contribute to the $t$-channel contribution \eq{PPfull}, 
Compton scattering cannot be dropped by invoking some collinearity argument.
In fact, our calculation reveals that Compton scattering not only contributes to a new constant, but more crucially to 
an additional logarithmic term, 
\beq
\left. -\frac{dE_{\gamma}}{dx} \right|^{v\to 1} = \frac{e^4 T^2}{96 \pi} 
\left[ \ln{\frac{4TE}{M^2}}  - \frac{5}{6} - \gamma + \frac{\zeta'(2)}{\zeta(2)} \right]  .
\label{loss-us}
\eeq
This logarithm is of collinear origin\footnote{Such 
potentially large logarithms were mentioned in Ref.~\cite{alam} in the case of the collisional energy loss of light partons.}, as is 
obvious from its divergence in the formal $M \to 0$ limit. 

In summary, if we aim to control the constant next to the leading logarithm in the $t$-channel contribution \eq{PPfull}, $s$ and $u$-channels must be included {\it for consistency}. This, in turn, brings an additional potentially large logarithm. 
For definiteness we state our complete result for the muon energy loss in the 
$v\to 1$ limit, obtained by adding \eq{loss-us} to \eq{PPfull},
\beqa
\left. -\frac{dE}{dx} \right|^{v\to 1} &=& \frac{e^4 T^2}{48 \pi} 
\left[ \ln{\frac{2 E}{e^2T}}  + \frac{1}{2} \ln{\frac{TE}{M^2}} +\ln{48} + \frac{3}{2} \left( \frac{\zeta'(2)}{\zeta(2)} -\gamma \right) - \frac{7}{6} \right] \nn \\
&\simeq& \frac{e^4 T^2}{48 \pi} \left[ \ln{\frac{2 E}{e^2T}}  + \frac{1}{2} \ln{\frac{TE}{M^2}} + 0.984 \right]  
\label{PPstu} \, .
\eeqa
This corrects the BT result \eq{BTfull}. 

Our paper is organized as follows. In section 2 we focus on the contribution $-dE_e/dx$ from $t$-channel exchange to the muon energy loss. In section 2.1 we present an exact relation between $-dE_e/dx$
and the muon self-energy. In section 2.2 we argue that the phase space can be conveniently decomposed
in terms of the Lorentz invariant momentum exchange $t$, into the regions $|t| < |t^\star|$ and 
$|t| > |t^\star|$, where the cut-off $t^\star$ satisfies $e^2T^2 \ll |t^\star| \ll T^2$, but otherwise is arbitrary. The contributions from the two regions are evaluated in sections 2.3 and 2.4 and summed in section 
2.5, where we derive the result \eq{PPfull} and discuss the incorrect assumption made in the BT calculation \cite{BT}. (For completeness, we repeat the calculation of \eq{PPfull} in Appendices B and C by following the procedure used in Ref.~\cite{BT}, \ie\ by decomposing the phase space into a soft and a hard domain with respect to $q^\star = |t^\star|^{1/2}$. We can thus precisely see where the approximation $q \ll E$ used in \cite{BT} fails when evaluating the hard contribution from $q > q^\star$.)
Section 3 is devoted to the contribution from Compton scattering $-dE_\gamma/dx$ quoted in \eq{loss-us}, and our results are summarized in section 4.

\section{Scattering off electrons}
\label{sec2}

\subsection{An exact relation: energy loss from self-energy}

Among the two processes contributing to the muon energy loss, namely scattering off thermal photons and scattering off thermal electrons (or positrons), we focus here on the latter mechanism, even though the idea of calculating the energy loss of a test particle from its self-energy is more general.
Our discussion below follows the BT calculation \cite{BT} of the soft contribution to the energy loss, 
which we generalize appropriately. 

\begin{figure}[b]
\bigskip
\centerline{\raisebox{-2mm}{\includegraphics[width=2.5cm]{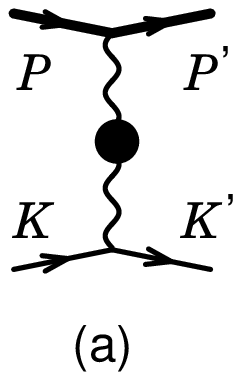}}
			\quad \quad
			\raisebox{-2mm}{\includegraphics[width=3.5cm]{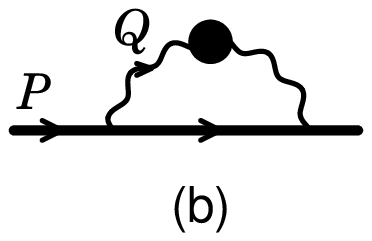}}}
\caption[*]{(a) $t$-channel scattering amplitude off electrons contributing to the muon interaction rate.
(b) The dressed muon self-energy.}
\label{fig:SE}
\end{figure}

Let us start by recalling that the collisional energy loss of a test particle is closely related to its interaction 
rate $\Gamma$, the latter being obtained from the imaginary part of the particle's self-energy $\Sigma$ evaluated at the energy 
$p_0 = E+i\epsilon$ \cite{Weldon:1983jn}. In the case of a muon,
\be
	\Gamma(E)
	=
	-\frac1{2E}\, \l( 1-n_F(E) \r)
	{\rm tr}\l[ (P\slash + M)\, {\rm Im}\,\Sigma(P)\r] \, .
\label{eq:Weldons formula}
\ee
For the $t$-channel interaction with electrons and positrons (involving a single photon exchange), the corresponding self-energy is the 1-loop graph depicted in Fig.~\ref{fig:SE}b, with a resummed photon propagator for reasons to be explained shortly.
Since the on-shell self-energy is gauge invariant, one may choose a convenient gauge in order to evaluate \eq{eq:Weldons formula}.
In Coulomb gauge, and without further approximation, the trace in \eq{eq:Weldons formula} reads \cite{BT}
\bea
\label{eq:tr}
    && {\rm tr}\l[ (P\slash + M)\, {\rm Im}\Sigma(P)\r]
	\;=\;
	-4\pi e^2 \l( 1+e^{-E/T} \r)
	\int_q \int_{-\infty}^\infty d\omega \l( 1+n_B(\omega) \r)
	\frac{{\cal A\, B}}{2E'} \, ,
	\\  && \quad
	{\cal A}
	\;=\;
	\rho_L(\omega,q)\l( 2E^2-E\omega-\bm{pq} \r)
	+2\rho_T(\omega,q)\l( p^2-E\omega+\bm{pq}-(\bm{pq}/q)^2 \r),
	\nonumber \\[1mm]  && \quad
	{\cal B}
	\;=\;
	(1-n_F(E'))\, \delta(E-E'-\omega) - n_F(E')\, \delta(E+E'-\omega) \, ,
	\nonumber
\eea
where $E' = \sqrt{(\bm p - \bm q)^2+M^2}$, $Q = (\omega, \bm q)$ is the photon momentum, and
$n_{B,F}$ denote the Bose-Einstein and Fermi-Dirac thermal distributions. We also use the shorthand notation
\be
\int_q  \equiv \int \frac{d^3 \bm q}{(2\pi)^3}
\eeq
and the spectral functions of the longitudinal and transverse photons \cite{BT}
\be
\rho_{L,T}(\omega,q) \equiv
-\frac1{\pi}\, {\rm Im}\l[ \Delta_{L,T}(\omega+i\epsilon,q) \r] \, ,
\label{eq:rho}
\ee
where $\Delta_{L,T}$ are the longitudinal and transverse photon propagators.

According to the assumption $M \gg T$, we have $n_F(E') \ll 1$ and the factor ${\cal B}$ in (\ref{eq:tr}) reduces to
$\delta(E-E'-\omega)$.
Performing the angular integral in (\ref{eq:tr}) yields
\bea
	\Gamma_e(E)
	&=&
	\frac{e^2}{2\pi v} \int_0^\infty dq\,q
	\int_{\omega_-}^{\omega_+} d\omega \l( 1+n_B(\omega) \r)
	\l\{ \rho_L(\omega,q) \l[ 1 - \l( \frac\omega{E} - \frac{t}{4E^2} \r) \r] \r.
	\nonumber \\
	&& \hskip 2cm
	+ \l.
	 \rho_T(\omega,q) \l[ v^2 - \frac{\omega^2}{q^2}
			+\l( \frac{t\,\omega}{E q^2} - \frac{t}{2E^2} - \frac{t^2}{(2Eq)^2} \r)
 	\r] \r\} .
\label{eq:Gamma}
\eea
Comparing to the analogous BT result, we observe that relaxing their assumption
$|\omega|, q \lsim T$ yields the additional terms put in between parentheses, and requires using the exact expression $\omega_\pm(q) = E - \sqrt{(p \mp q)^2+M^2}$ for the bounds instead of the approximation $\pm vq$.

The energy loss per unit length is then obtained by weighting the differential interaction rate by $\omega/v$,
\bea
	-\frac{dE_e}{dx}
	&=&
	\frac{e^2}{2\pi v^2} \int_0^\infty dq\,q
	\int_{\omega_-}^{\omega_+} d\omega\, \omega \l( 1+n_B(\omega) \r)
	\l\{ \rho_L(\omega,q) \l[ 1 - \l( \frac\omega{E} - \frac{t}{4E^2} \r) \r] \r.
	\nonumber \\
	&& \hskip 2cm
	+ \l.
	 \rho_T(\omega,q) \l[ v^2 - \frac{\omega^2}{q^2}
			+\l( \frac{t\,\omega}{E q^2} - \frac{t}{2E^2} - \frac{t^2}{(2Eq)^2} \r)
 	\r] \r\} .
\label{eq:dEdx_full}
\eea
As discussed in Ref.~\cite{BT}, the additional factor of $\omega$ is improving the infrared behavior of the integral (\ref{eq:dEdx_full}) compared to (\ref{eq:Gamma}). In fact, evaluating $\Gamma_e$ from Eq.~(\ref{eq:Gamma}) with spectral functions obtained as the discontinuity of the 1-loop resummed propagators $\Delta^{\rm 1-loop} = (\Delta_0^{-1} - \Pi^{\rm 1-loop})^{-1}$, yields an infrared divergence due to soft transverse exchanges.
In contrast, weighting the integrand of $\Gamma_e$ by $\omega$ leads to a well-defined energy loss \eq{eq:dEdx_full}.
%SP when using the 1-loop resummed photon spectral function.

\subsection{Phase space decomposition}
\label{sec:PS decomposition}

The necessity of expressing the energy loss in terms of a {\it dressed} exchanged photon propagator (obtained by resumming the 1-loop photon self-energy) arises from the long-range nature of the Coulomb interaction. 
However, this infrared dynamics becomes unimportant for short-range interactions.
This motivates, on physical grounds, a decomposition of the phase space in Eq.~(\ref{eq:dEdx_full}).

\begin{figure}[ht]
\centerline{\includegraphics[width=5cm]{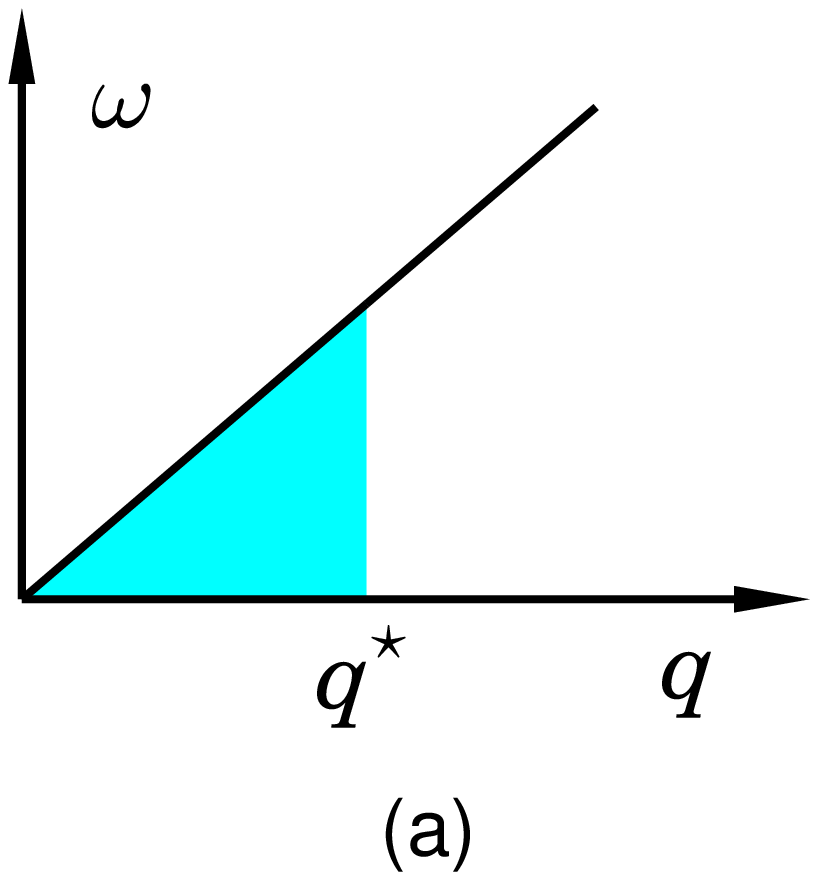}\quad\quad
			\includegraphics[width=5cm]{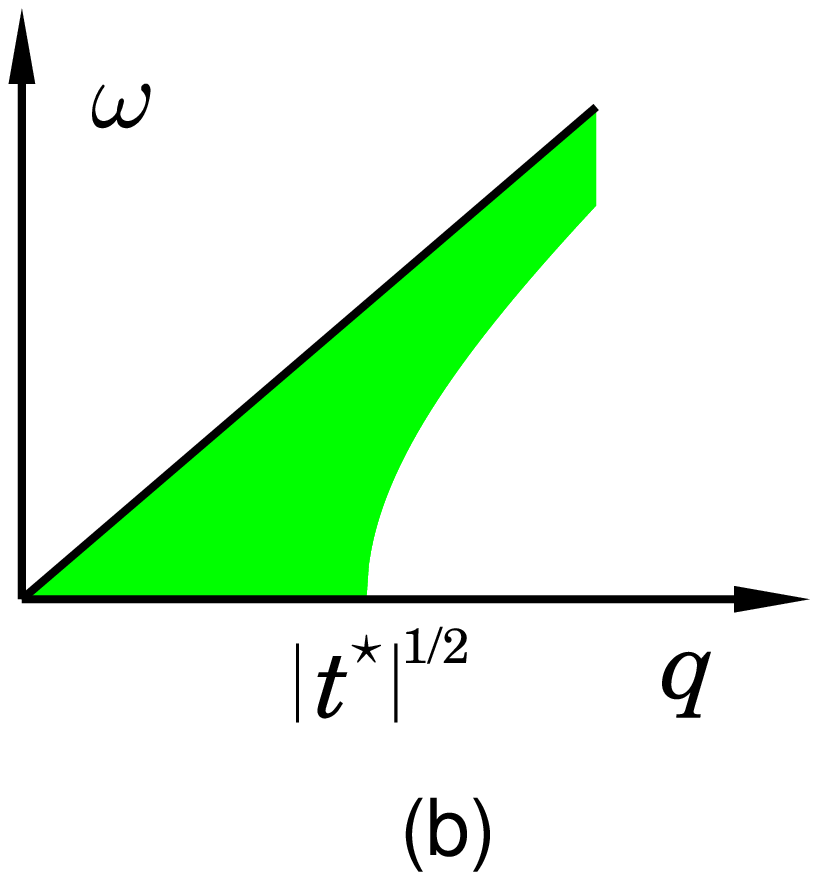}}
\caption[*]{Two ways of decomposing the exchanged photon phase space. (a) Braaten-Yuan prescription,
by introducing a cut-off $q^\star$ separating a soft region (shaded) from a hard one, as applied in \cite{BT}. 
(b) Our approach, using a cut-off $t^\star \equiv -(q^\star)^2$ in the invariant momentum transfer.}
\label{fig:decompose phase space}
\end{figure}

In the BT calculation of the energy loss, the matching procedure developed by Braaten and Yuan 
\cite{Braaten:1991dd} is used. A momentum scale $q^\star$, chosen as $e T \ll q^\star \ll T$ but otherwise arbitrary, is introduced to separate soft interactions, with $q < q^\star$, from hard ones  with $q > q^\star$, see Fig.~\ref{fig:decompose phase space}a.
The soft contribution to $dE/dx$ is evaluated using the `hard thermal loop' (HTL) approximation \cite{pisarski,BI} for the dressed photon propagator.
On the other hand, the hard $q > q^\star$ contribution in (\ref{eq:dEdx_full}) is obtained by keeping only the leading term in the expansion $\rho \propto {\rm Im}\l[ (\Delta_0^{-1} - \Pi)^{-1} \r] = \Delta_0^2\, {\rm Im}\l[ \Pi \r] + {\cal O}(\Pi^2)$. This approximation corresponds to evaluating the elastic scattering amplitude in Fig.~\ref{fig:SE}a with the tree-level photon propagator
(which would yield an infrared divergent result for the energy loss in the absence of the cut-off $q^\star$). 
The sum of the soft and hard contributions should, of course, be independent of the arbitrary scale $q^\star$, as was verified in \cite{BT}.
However, this consistency check could not reveal that BT's result for the hard contribution is incomplete beyond logarithmic accuracy, as we will show in the following.

For this purpose it will be convenient to decompose the phase space with respect to a cut-off $t^\star \equiv -(q^\star)^2$ in the invariant momentum transfer, as illustrated in Fig.~\ref{fig:decompose phase space}b.
This choice is motivated by two facts. 
First, in the region where $|t| < |t^\star|$, the HTL approximation for the photon propagator is known to be valid \cite{Peshier:1998dy} although $\omega$ and $q$ can be individually large. In fact, while the HTL approximation is usually derived under the assumption $\omega,q \ll T$ (implying $\omega^2+q^2 \ll T^2$), it actually holds if the Minkowski `norm' 
$|\omega^2-q^2|$ is small compared to $T^2$ \cite{Peshier:1998dy}. 
Secondly, the calculation of the contribution where $|t| > |t^\star|$ is more transparent (see section \ref{sec:tchannelhard}), 
since in this region the squared scattering amplitude is a function of the Mandelstam invariants only. 
We stress that when $|t| > |t^\star|$ 
we can indeed neglect the `medium modifications' to the matrix elements since the region $|t| > |t^\star|$ is
contained in the hard $q > q^\star$ Braaten-Yuan region.

\subsection{Contribution from $|t| < |t^\star|$}
\label{sec:tchannelsoft}

We calculate this contribution to $dE_e/dx$ from \eq{eq:dEdx_full} by changing variables to $t = \omega^2-q^2$ and $x = \omega/q$. Since $|t| < |t^\star|$, we can omit the terms of order ${\cal O}(|t|^{1/2}/E)$ and approximate the bounds on $x$ by $\pm v$. The term $\sim \omega/E$ in the integrand of \eq{eq:dEdx_full} is easily checked to be exponentially suppressed. Using the HTL approximation for the spectral function we obtain
\beq
-\l. \frac{dE_e}{dx} \r|_{|t| < |t^\star|}
	=
	\frac{e^2}{4\pi v^2}
	\int_{t^\star}^0 dt\, (-t)
	\int_{-v}^{v} dx\, \frac{x}{(1-x^2)^2}
	\l( 1+n_B(\omega) \r)
	\l[ \rho_L + (v^2 - x^2) \rho_T \r] , 
\eeq
where $\omega = x \sqrt{-t/(1-x^2)}$. From a simple parity argument, we can replace the factor $1+n_B(\omega)$ by its 
even part, \ie\ $1+n_B(\omega) \rightarrow \frac12$, and we find
\beq
-\l. \frac{dE_e}{dx} \r|_{|t| < |t^\star|}
	=
	\frac{e^2}{8\pi v^2}
	\int_{-v}^{v} dx\, \frac{x}{(1-x^2)^2}
	\int_{t^\star}^0 dt\, (-t)
	\l[ \rho_L + (v^2 - x^2) \rho_T \r] .
\label{tsoft2}
\eeq
Using the HTL longitudinal and transverse photon propagators
\beq	
\Delta_L(\omega, q) = \frac{1}{q^2+\Pi_L(x)}  \, , \quad 
\Delta_T(\omega, q) = \frac{1}{\omega^2-q^2-\Pi_T(x)} 
\label{propagators}
\eeq
with the self-energies \cite{BI}
\beqa
\Pi_L(x) &=& m_D^2 \l[ 1-Q(x) \r] \, , \nn \\
\Pi_T(x) &=& \frac{m_D^2}{2} \l[ x^2+(1-x^2)Q(x) \r] 
		  = \frac{m_D^2}{2}\, x (1-x^2) Q'(x) \, , \label{self} \\
Q(x) &\equiv& \frac{x}{2}\ln\frac{x+1}{x-1} \, ,\nn 
\eeqa
where $m_D = eT/\sqrt{3}$ is the Debye mass in the QED plasma, the expression \eq{tsoft2} becomes
\beqa
-\l. \frac{dE_e}{dx} \r|_{|t| < |t^\star|} &=& \frac{e^2}{8\pi^2 v^2} \int_{-v}^{v} dx\, \frac{x}{1-x^2}
\int^{t^\star}_0 dt\, {\rm Im} \l[ \frac{t}{t-{\tilde \Pi}_L(x)} -\frac{v^2 - x^2}{1-x^2} \frac{t}{t-\Pi_T(x)}\r]  \nn \\
&=& \frac{e^2}{8\pi^2 v^2} \int_{-v}^{v} dx\, \frac{x}{1-x^2} {\rm Im} 
\l[ {\tilde \Pi}_L \ln{\frac{|t^\star|+{\tilde \Pi}_L}{{\tilde \Pi}_L}} 
-\frac{v^2 - x^2}{1-x^2} \Pi_T \ln{\frac{|t^\star|+\Pi_T}{\Pi_T}}  \r] \, . \nn \\
\label{tsoft3}
\eeqa
We used \eq{eq:rho} and introduced ${\tilde \Pi}_L(x) \equiv (1-x^2)\Pi_L(x)$. 
We now take advantage of $|t^\star| \gg |{\tilde \Pi}_L(x)|, |\Pi_T(x)|$, and extract the $|t^\star|$-dependence, writing for instance
\beq
\ln{\frac{|t^\star|}{\Pi_T}} = \ln{\frac{2|t^\star|}{m_D^2}} + \ln{\frac{m_D^2}{2 \Pi_T}} \, .
\eeq
We arrive at 
\beqa
-\l. \frac{dE_e}{dx} \r|_{|t| < |t^\star|} &=& \frac{e^4T^2}{48\pi v} \l[ 1 - \frac{1-v^2}{2v}\,\ln\frac{1+v}{1-v} \r] 
\ln\l( \frac{2 |t^\star|}{m_D^2} \r) \nn \\
&-& \frac{e^4T^2}{48\pi^2 v^2} \int_{-v}^{v} dx\, \frac{x}{1-x^2} {\rm Im} 
\l[ f_L \ln f_L - \frac{v^2-x^2}{1-x^2} f_T \ln f_T \r] \, ,
\label{tsoft4}
\eeqa
where 
\beqa
f_L(x) &=& 2 \, {\tilde \Pi}_L(x) /m_D^2 = 2 (1-x^2)\l( 1-Q(x) \r) , \nn \\
f_T(x) &=& 2 \, \Pi_T(x) /m_D^2 = x (1-x^2) Q'(x) \, .
\eeqa
The integral in \eq{tsoft4} seems difficult to evaluate analytically for arbitrary $v$. However, in the limit $v \to 1$ we are interested in, the result is surprisingly simple: the integral vanishes, as can be checked numerically. Hence,
\be
-\l. \frac{dE_e}{dx} \r|_{|t| < |t^\star|}^{v \to 1} = \frac{e^4T^2}{48\pi}\, \ln{\frac{6 |t^\star|}{e^2 T^2}} \, .
\label{softtstar}
\ee
We end this section by presenting an alternative and fully analytical way to obtain the result \eq{softtstar} 
-- which proves indirectly that the integral in \eq{tsoft4} indeed vanishes for $v = 1$.
Referring to Fig.~\ref{fig:decompose phase space}, the kinematic region $|t| < |t^\star|$ is obviously given by the reunion of the soft region
$q^2 < (q^\star)^2 = |t^\star|$ and the region where $q^2 > |t^\star|$ and $|t| < |t^\star|$, or
equivalently  $(1-x^2)|t^\star| < |t| < |t^\star|$, since $|t| = (1-x^2)q^2$. The contribution 
from the latter region is easy to calculate along the lines which led to \eq{tsoft3}. When $v=1$ we get
\beqa
-\l. \frac{dE_e}{dx} \r|_{|t| < |t^\star|,\, q > q^\star} &=& \frac{e^2}{8\pi^2} \int_{-1}^{1} dx\, \frac{x}{1-x^2}
\int^{t^\star}_{(1-x^2)t^\star} dt\, {\rm Im} \l[ \frac{{\tilde \Pi}_L}{t - {\tilde \Pi}_L}- \frac{\Pi_T}{t - \Pi_T}  \r]  \nn \\
&=& \frac{e^4 T^2}{48\pi} \int_{-1}^{1} dx\, \frac{3x^2}{2} \ln{\left( \frac{1}{1-x^2} \right)} 
= \frac{e^4 T^2}{48\pi}  \l[ \frac{8}{3} - \ln 4 \r] \, ,
\label{interregion}
\eeqa
where we again used $|t^\star| \gg |{\tilde \Pi}_L(x)|, |\Pi_T(x)|$. The contribution from 
$q^2 < |t^\star|$ is precisely the BT `soft' contribution, which reads for $v=1$ \cite{BT}
\beq
-\l. \frac{dE_e}{dx} \r|_{q < q^\star}= \frac{e^4 T^2}{24 \pi}\left[ \ln{\frac{q^\star}{e T}} + 0.256 \right]  \, .
\label{BTsofta}
\eeq
In Appendix B we rederive this result and determine the constant analytically,
\beq
-\l. \frac{dE_e}{dx} \r|_{q < q^\star}=  \frac{e^4 T^2}{24 \pi}\left[ \ln{\frac{q^\star}{e T}} +\frac{\ln 24}{2} -\frac{4}{3} \right] \, .
\label{PPsoftb}
\eeq
By adding \eq{interregion} and \eq{PPsoftb} we confirm the result \eq{softtstar} for the contribution 
from the domain $|t| < |t^\star|$.

\subsection{Contribution from $|t| > |t^\star|$}
\label{sec:tchannelhard}

As already argued in section \ref{sec:PS decomposition}, when $|t| > |t^\star|$ thermal corrections to the exchanged photon propagator can be ignored. In this kinematic domain the energy loss can thus be obtained from the general relation \cite{BT}
\beqa
  - \frac{dE_i}{dx}
  &=&
  \frac{1}{2Ev}
	\int_k \frac{n_i(k)}{2k}
	\int_{k'} \frac{\bar n_i(k')}{2k'}\,
	\int_{p'} \frac1{2E'}\,	(2\pi)^4\delta^{(4)}(P+K-P'-K')\,
	\frac{1}{d}\sum_{\rm spins}\l| {\cal M}_i\r|^2\, \omega  \nn \\
\label{eq:dEdx0}
\eeqa
by inserting a factor $\Theta(|t| - |t^\star|)$ in the integrand. The tree-level matrix element ${\cal M}_i$ describes the scattering off a target particle of type $i$.
Summing $|{\cal M}_i|^2$ over initial and final spin states and dividing by the degeneracy factor $d$ of the incoming test particle gives its energy loss $- dE_i/dx$ averaged over spin states.
Furthermore, $n_i(k) = (\exp(k/T) \pm 1)^{-1}$ is the thermal distribution of the target particles, and $\bar n_i = 1 \pm n_i$ accounts for the Bose enhancement or Pauli blocking for the scattered state.
In line with our previous considerations, the target particles are assumed to be massless.

The tree-level matrix elements squared depend only on the Mandelstam in\-va\-riants $s$ and $t$.
Then, as derived in Appendix A, the phase space integral in \eq{eq:dEdx0} can be reduced to
\beq
 - \frac{dE_i}{dx}
  = \frac{1}{v}\, d_i
	\int_k \frac{n_i(k)}{2k}
	\l(	1 - \frac{s+M^2}{s-M^2}\frac{k}E \r)
	\int_{t_{\rm min}}^0 dt\, (-t)\, \frac{d\sigma_i}{dt} \, ,
\label{eq:dEdx}
\eeq
where only the approximation $\bar n_i \rightarrow 1$ has been made, which is also justified in Appendix A.
Apart from this simplification, the expression \eq{eq:dEdx} is exact, in particular with respect to the kinematics of the scattering process.
The bound in the $t$-integral is
\beq
	t_{\rm min} = -\frac{(s-M^2)^2}s \, ,
\eeq
and we introduced the differential cross section
\beq
	\frac{d\sigma_i}{dt}
	=
	\frac1{16\pi (s-M^2)^2}\,
	\frac1{d\, d_i} \sum_{\rm spins}\l| {\cal M}_i \r|^2  \ \ ,
\label{crossdef}
\eeq
where $d_i$ is the spin degeneracy of the target particles.

Our expression \eq{eq:dEdx} generalizes a formula used by Bjorken \cite{bj} for a massless test particle of infinite energy, to the massive and finite-energy case. It will allow us to calculate in a rather compact way the contribution from $|t| > |t^\star|$ to the collisional energy loss. 
Focusing on the large $E$ limit and inserting $\Theta(|t| - |t^\star|)$ 
in the integrand of \eq{eq:dEdx} we obtain
\beq
   \l. -\frac{dE_i}{dx}\r|_{|t| > |t^\star|}^{v \to 1}  =  d_i
	\int_k \frac{n_i(k)}{2k}
	\int_{t_{\rm min}}^{t^\star} dt\, (-t)\, \frac{d\sigma_i}{dt} \, .
\label{eq:dEdx_asy}
\eeq
The energy loss $-dE_e/dx$ arising from $|t| > |t^\star|$ of a muon scattering off electrons and positrons\footnote{The contribution of positrons 
is identical to that of electrons for an $e^\pm \gamma$ plasma with vanishing chemical potential, and is accounted for by a factor $2$ in Eq.~\eq{tampsquared}.} is obtained from \eq{eq:dEdx_asy} with \eq{crossdef} and
\beq 
2 \sum_{\rm spins} \l| {\cal M}_{e^-} \r|^2 
%% = 2 . 8 e^4 \frac{\tilde s^2 + \tilde u^2 + 2M^2t}{t^2} 
= 32\,e^4 \left[ \frac{(s-M^2)^2}{t^2} + \frac{s}{t} + \frac{1}{2} \right]  .
\label{tampsquared}
\eeq
We mention that the two last terms of \eq{tampsquared} contribute to the $t$-integral in \eq{eq:dEdx_asy} as a constant,
\beq
\frac{1}{(s-M^2)^2} \int_{t_{\rm min}}^{t^\star} dt\, (-t) \, \left[ \frac{s}{t} + \frac{1}{2} \right] \simeq - \frac{3}{4} \ \ ,
\eeq
where we used $s \gg M^2 \gg |t^\star|$. This results in a non-zero contribution to $-dE_e/dx$,
\beq
\frac{e^4 T^2}{48 \pi} \cdot \left(-\frac{3}{4} \right) \ \ ,
\label{lastterms}
\eeq
which arises from the very hard region $-t \sim -t_{\rm min} \sim s$, \ie\ $q \sim \morder{E}$.
This proves that the assumption $q \ll E$ used in \cite{BT} is inappropriate when calculating $-dE/dx$ beyond logarithmic accuracy.

Only the first term of \eq{tampsquared} is sensitive to the cut-off $t^{\star}$, since for $t^{\star} \to 0$ it yields a logarithmic divergence in the infrared,
\beq
\frac{1}{(s-M^2)^2} \int_{t_{\rm min}}^{t^\star} dt\, (-t) \, \left[ \frac{(s-M^2)^2}{t^2} \right] = \ln \frac{|t_{\rm min}|}{|t^\star|}
\simeq \ln \frac{s}{|t^\star|} \ \ .
\label{reg}
\eeq
With the help of the definite integrals
\beqa
\int_0^\infty dx \, n_F(x) \, x  &=& \frac{\pi^2}{12} \, , \\
\int_0^\infty dx \, n_F(x) \, x \ln{x} &=& \frac{\pi^2}{12} \left[ 1- \gamma + \ln{2} + \frac{\zeta'(2)}{\zeta(2)}\right] \, ,
\eeqa
the contribution of the first term of \eq{tampsquared} to \eq{eq:dEdx_asy} is found to be
\beq
\frac{e^4 T^2}{48 \pi} 
\left[ \ln\frac{8TE}{|t^\star|} - \gamma + \frac{\zeta'(2)}{\zeta(2)} \right] .
\label{firstterm}
\eeq
Adding the contributions \eq{lastterms} and \eq{firstterm} we obtain
\beq
- \left. \frac{dE_e}{dx} \right|_{|t|>|t^\star|}^{v \to 1} = \frac{e^4T^2}{48\pi} \l[ \ln\frac{8ET}{|t^\star|} - \gamma - \frac{3}{4} +\frac{\zeta'(2)}{\zeta(2)} \r]  .
\label{hardtstar}
\eeq

\subsection{Complete $t$-channel result and discussion}

Combining the contributions from $|t|<|t^\star|$ and $|t|>|t^\star|$ given by \eq{softtstar} and \eq{hardtstar}, 
we find the muon energy loss from scattering off electrons and positrons as
\beq
\left. -\frac{dE_e}{dx} \right|^{v\to 1} = \frac{e^4 T^2}{48 \pi} 
\left[ \ln{\frac{2 E}{e^2T}}  + \ln{24}  - \gamma + \frac{\zeta'(2)}{\zeta(2)} - \frac{3}{4} \right] ,
\label{PPfull2} 
\eeq
as already quoted in \eq{PPfull}. 

We stress that the leading logarithmic term in \eq{PPfull2} arises from the 
region $m_D^2 \ll -t \ll s \sim ET$. Since $-t = - (K-K')^2 = 2 k k' (1 -\cos{\theta_{kk'}})$ and
$k \sim T$, this implies $k' = |{\bm k} + {\bm q}|\ll E$. The approximation 
$q \ll E$ used in \cite{BT} is thus legitimate at logarithmic accuracy. Beyond logarithmic accuracy however, the region
contributing to \eq{PPfull2} extends to $m_D^2 \lsim -t \lsim s \sim ET$. In particular 
the region of maximal transfers $-t \sim -t_{\rm min} \sim  s \sim ET$ 
(\ie, $k' \sim E$) affects the constant. Thus the latter could
not be correctly determined in \cite{BT}, where the approximation $q \ll E$ was used. 
As mentioned in the Introduction (see \eq{sestimate} and \eq{angles}), $k' \sim E$ also implies that the angular domain $\theta_{p'k'} \sim \sqrt{T/E} \ll 1$ contributes to the constant.

The constant in our result \eq{PPfull2} differs from that of the BT result \eq{BTfull}. 
In order to confirm our result, we present an alternative calculation in Appendices B and C. There we follow the BT approach by using the familiar Braaten-Yuan decomposition of the phase space into a soft $q<q^\star$ and a hard $q>q^\star$ kinematic domain, see Fig.~\ref{fig:decompose phase space}a.
In Appendix B we confirm (see \eq{soft4}) the BT result \eq{BTsofta} for the soft contribution in the $v\to 1$ limit.
In Appendix C, we repeat the BT calculation of the hard $q > q^\star$ contribution {\it without using the approximation $q \ll E$}. As expected, we find that the constant is sensitive to the hard domain $q \sim E$.
As shown explicitly in Appendix C, the approximation $q \ll E$ would amount to neglect terms both in $\l| {\cal M}_e \r|^2 $ and in the $\delta$-function for energy conservation, which are important to determine the constant. Those corrections lead to the result \eq{totalhard}, instead of \eq{BThard} as found by BT\footnote{It is a coincidence that the last term $\sim - 3/4$
of \eq{totalhard} missed by BT is identical to the contribution \eq{lastterms}. Hence, in effect the BT result is the same as what we would obtain by keeping only the term $\sim (s-M^2)^2/t^2$ in \eq{tampsquared}.}.

Adding the soft \eq{soft4} and hard \eq{totalhard} contributions we recover \eq{PPfull2}, found within our decomposition of phase space using an invariant separation scale $t^\star$. 
We view this as a corroborating evidence of the correctness of our results.

\section{Compton scattering}
\label{sec:Compton}

\begin{figure}[ht]
\centerline{\raisebox{-5mm}{\includegraphics[width=7cm]{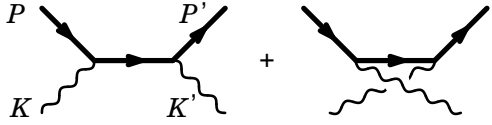}}}
\caption[*]{Amplitude ${\cal M}_\gamma$ for Compton scattering.}
\label{fig:compton}
\end{figure}

The contribution $-dE_{\gamma}/dx$ to the muon energy loss from Compton scattering (see Fig.~\ref{fig:compton}) is dominated, as we will shortly see, by hard transfers $-t \sim s$. Thus it can be obtained from \eq{eq:dEdx_asy} by setting $t^\star =0$ and by using
\beqa
\sum_{\rm spins}\l| {\cal M}_\gamma \r|^2 &=& 8e^4 \l[ \l( \frac{-\tilde u}{\tilde s}+2 M^2\frac{\tilde s+2M^2}{\tilde s^2} \r)
+\l( \frac{\tilde s}{-\tilde u}+2 M^2\frac{\tilde u+2M^2}{\tilde u^2} \r) \r.
\nn \\ && \hskip 5.8cm
+ \l. 2 M^2\frac{\tilde s+\tilde u+4M^2}{\tilde s \tilde u} \r] \, ,
\label{eq:M2Compton}
\eeqa
where we define $\tilde u \equiv u-M^2$ and $\tilde s \equiv s-M^2$, which satisfy $\tilde s + \tilde u + t = 0$. The three terms in \eq{eq:M2Compton} correspond to the contributions from the $s$ and $u$-channels, and from the interference term.
Using \eq{eq:M2Compton} in our formula \eq{eq:dEdx_asy} we can easily show that all terms which explicitly depend on $M^2$ in \eq{eq:M2Compton} yield contributions to $-dE_{\gamma}/dx$ which are suppressed by at least one power of $s \sim E T$ when $E \to \infty$. Thus for our purposes \eq{eq:M2Compton} can be approximated by
\beq
\sum_{\rm spins}\l| {\cal M}_\gamma \r|^2 \simeq 8e^4 \l[ \frac{-\tilde u}{\tilde s} +\frac{\tilde s}{-\tilde u} \r]  \, .
\label{eq:M2Comptonappr}
\eeq
This yields the integral
\beqa
\int_{t_{\rm min}}^0 dt\, (-t)\, \frac{d\sigma_\gamma}{dt} &=& \frac{e^4}{8\pi}
\int_{\tilde u_{\rm min}}^{\tilde u_{\rm max}} d\tilde u\, \frac{\tilde u + \tilde s}{\tilde s^2}
\l[ \frac{-\tilde u}{\tilde s} +\frac{\tilde s}{-\tilde u} \r] + \morder{1/s}
\label{sigmagamma0} \\
&=& \frac{e^4}{8\pi} \l( \ln\frac{s}{M^2} - \frac56 \r) + \morder{1/s} \, ,
\label{sigmagamma}
\eeqa
where we changed variables from $t$ to $\tilde u$, with the bounds $\tilde u_{\rm min} = -\tilde s$, $\tilde u_{\rm max} = -M^2\, \tilde s/s$.

The leading logarithm arises from the $u$-channel, more specifically from the kinematic region $\tilde u_{\rm min} \ll \tilde u \ll \tilde u_{\rm max}$, \ie\ $M^2 \ll -\tilde u \ll s$ when $s \gg M^2$. Since in this region $-t \simeq \tilde s \simeq s$ (recall that $\tilde s + \tilde u + t = 0$), the physical interpretation of the logarithmic enhancement is the same as for the total cross section $\sigma_\gamma = \int dt\, d\sigma_\gamma/dt$. The latter behaves in the high energy limit as $\alpha^2 s^{-1} \ln(s/M^2)$ \cite{PS}, with the logarithm originating from backward scattering of the photons in the center of momentum frame.
We thus infer that the Compton contribution to the energy loss is dominated, at large energies and to leading logarithmic accuracy, by the same mechanism.

Plugging \eq{sigmagamma} into \eq{eq:dEdx_asy}, and using
\beqa
\int_0^\infty dx \, n_B(x) \, x  &=& \frac{\pi^2}{6} \, , \\
\int_0^\infty dx \, n_B(x) \, x \ln{x} &=& \frac{\pi^2}{6} \left[ 1- \gamma + \frac{\zeta'(2)}{\zeta(2)}\right]
\eeqa
for the integral over $k$, we obtain
\beq
\left. -\frac{dE_{\gamma}}{dx} \right|^{v\to 1} = \frac{e^4 T^2}{96 \pi}
\left[ \ln{\frac{4 T E}{M^2}} - \frac{5}{6} - \gamma + \frac{\zeta'(2)}{\zeta(2)} \right] \, ,
\label{usfull}
\eeq
as quoted in \eq{loss-us}.

The logarithm in \eq{usfull} arises from $M^2 \ll -\tilde u \ll s$, while
the regions $-\tilde u  \sim M^2$ and $-\tilde u  \sim s$ only contribute to the constant.
Hence the complete expression \eq{usfull} stems from\footnote{Since $M \gg T$, the contribution from $s$ and $u$-channels arises from exchanges where thermal corrections to the muon propagator are suppressed. The calculation using a bare muon propagator in Fig.~\ref{fig:compton} is thus legitimate.} $M^2 \lsim -\tilde u \lsim s$, including also very hard exchanges $-t \sim s \sim ET \gg M^2$.
>From $-t = - (K-K')^2 = 2 k k' (1 -\cos{\theta_{kk'}}) \sim s \sim ET$, the typical values of $k' = k +\omega$ contributing to \eq{usfull} are $k' \sim k'_{\rm max} \sim E$. From \eq{sestimate} we thus find that $s$ and $u$-channel contributions arise from 
the angular domain $\theta_{p'k'} \sim \sqrt{T/E} \ll 1$, as anticipated in the Introduction.  

It is instructive to write the integral $\sim \int d\tilde u /\tilde u$ appearing in \eq{sigmagamma0} as an integral over the angle between ${\bm p}$ and ${\bm k}'$ by using
\beq
\tilde u = (P-K')^2 -M^2 = -2 P K' \mathop{\simeq}_{v\to 1} -2E k' (1 -\cos{\theta_{pk'}}) \, .
\label{uestimate}
\eeq
Hence, in order to satisfy 
$-\tilde u \lsim s \sim ET$, we must have $\theta_{pk'} \ll 1$ in \eq{uestimate}, and the domain $M^2 \lsim -\tilde u \lsim s$ leading to the logarithm translates into $M/E \lsim \theta_{pk'} \lsim \sqrt{T/E}$. 
Consequently, the result \eq{usfull} stems from the angular regions
\beq
M/E \lsim \theta_{pk'} \lsim \sqrt{T/E} \, , \quad
 \theta_{p'k'} \sim \sqrt{T/E} \, .
\label{suangular}
\eeq

\section{Summary and outlook}

In this study we have reconsidered the Braaten-Thoma calculation \cite{BT} of the muon collisional energy loss in a hot QED plasma. For the $t$-channel contribution (scattering off electrons and positrons), we have shown that controlling the constant next to the leading logarithm requires, in the limit $E \gg M^2/T$, considering the region where the invariant transfer $-t$ is on the order of its maximal value, $-t \sim -t_{\rm min} \simeq s \gg M^2$.
The BT calculation of the $t$-channel contribution, which was based on the invalid assumption $q \ll E$, evaluated the constant incorrectly.
We obtained the corrected result for this contribution in Eq.~\eq{PPfull}.

We showed that the `constant' is sensitive to the angular domain
$\theta_{p'k'} \sim \sqrt{T/E}$, \ie\ to collinear configurations when $E \to \infty$.
Thus, for consistency the contribution from $s$ and $u$-channels (Compton scattering), which arises from similar configurations, {\it must} be included in the energy loss.
As already stressed in the Introduction, this is our main message. Removing Compton scattering from the definition of energy loss implies that we have to give up determining the constant next to the leading logarithm in the $t$-channel contribution. In other words, working beyond logarithmic accuracy is meaningful only with Compton scattering taken into account.
The Compton process yields a potentially large `collinear' logarithm $\propto \ln(ET/M^2)$, see Eq.~\eq{loss-us}.
It arises from hard transfers $-t \simeq s$, and was previously neglected in \cite{BT}.

It will be interesting to study the consequences of our findings for the collisional energy loss of a heavy quark in a hot QCD plasma, and to see how the results of \cite{BTqcd} are modified\footnote{A first step in this direction was done in \cite{Peshier:2006hi}.}.
We also point to the necessity, in phenomenological studies, to take into account the finite experimental angular resolution $\delta$. If the latter is of the order of
$\sqrt{T/E}$ (or larger), the final state configurations with $\theta_{p'k'} < \delta$ should be removed from
the definition of energy loss, modifying our full result \eq{PPstu} by introducing a $\delta$-dependence.
Since (quasi)-collinear configurations $\theta_{p'k'} \sim \sqrt{T/E}$ correspond to hard exchanges
$q \sim E$, the angular resolution $\delta$ will actually translate to an upper cut-off in $q$,
above which the elastic processes under consideration will not contribute to an observable energy loss.

%%\acknowledgments

\appendix

\section{Thermal phase space}

For a given function $f(s,t,\omega)$ depending on the Mandelstam invariants and the energy transfer $\omega=E-E'$, we calculate
the functional
\be
	{\cal I}[f]
	=
	\frac1{2E}\int_k \frac{n(k)}{2k}
	\int_{k'} \frac{\bar n(k')}{2k'}
	\int_{p'} \frac1{2E'}\,
	(2\pi)^4 \delta^{(4)}(P+K-P'-K')\, f(s,t,\omega) \, .
\label{eq:Idef}
\ee

We start with the $k'$-integral, for which the specific form of $f(s,t,\omega)$ is not relevant, since
$s$ and $t$ are determined by $\bm k$, $\bm p'$ and $\bm p$ only. Following a standard procedure we write
\[
	\int \frac{d^3 {\bm k}'}{(2\pi)^3}\, \frac1{2k'}
	=
	2\pi \int \frac{d^4k'}{(2\pi)^4}\, \Theta(k_0')\, \delta( K'^2) \, ,
\]
and evaluate
\beq
	\int_{k'} \frac{\bar n(k')}{2k'}\, (2\pi)^4 \delta^{(4)}(P+K-P'-K')
	=
	2\pi \bar n(\underline{k_0'}) \Theta(\underline{k_0'})\,
	\delta\l( \underline{K}'^2 \r) .
\label{eq:k'-int}
\eeq
Here $\underline{K}' = K+P-P'$ is fixed by momentum conservation. 
With $P-P' = Q = (\omega, \bm q)$, we have in particular $\underline{k_0'} = k+\omega$.

In order to proceed with the $p'$-integral, we specify a coordinate system.
We choose the $z$-axis along the direction of $\bm p$, and orient the $yz$-plane to contain $\bm k$,
\beqa
	\bm p &=& (0,0,1)p \, , \nonumber \\
	\bm k &=& (0,\sin\psi,\cos\psi)k \, , \nonumber \\
	\bm p' &=& (\sin\theta\sin\phi,\sin\theta\cos\phi,\cos\theta)p' \, .
\label{eq:angles}
\eeqa
The integral over the azimuthal angle $\phi$ is readily performed with the help of the $\delta$-function 
in \eq{eq:k'-int}. We first express its argument in terms of the Mandelstam invariants $t=Q^2$ and $s=(K+P)^2=M^2+2KP$,
\beqa
\underline{K}'^2 &=& (K+Q)^2 = 2KQ+Q^2 = 2K(P-P')+t = s-M^2 + t - 2KP' \, .
\eeqa
Writing $KP' = kE'- \bm k \bm p'$ and using \eq{eq:angles}, we find $\underline{K}'^2 = A+B\cos\phi$, with
\beqa
A &=& s -M^2 +t-2kE'+2kp'\cos\psi\cos\theta \, , \nn  \\
B &=& 2kp'\sin\psi\sin\theta \, .
\label{AB}
\eeqa
Consequently,
\beq
\int_0^{2\pi} d\phi\, \delta\l( \underline{K}'^2 \r) = \frac2{\sqrt{g}}\, \Theta(g) \, ,
\label{eq:phi-int}
\eeq
with $g = B^2-A^2$. The $\Theta$-function reflects the kinematic constraints imposed by energy-momentum conservation.

For the remaining integrals we change variables, from $p'$ and $\cos\theta$ to
\beqa
t &=& 2\l(M^2 - EE' + p p' \cos\theta \r), \nonumber \\
\omega &=& E - E' \, ,
\label{eq:change_variables}
\eeqa
with the Jacobian $E'/(2p\,p'^2)$. 
Using \eq{eq:k'-int}, \eq{eq:phi-int}, the expression \eq{eq:Idef} becomes
\beq
{\cal I}[f] = \frac1{16\pi^2 pE} \int_k \frac{n(k)}{2k} \int_{-\infty}^0 dt 
\int_{-\infty}^{\infty} d\omega \frac{\Theta(g)}{\sqrt{g}}\, \bar{n}(k+\omega) \Theta(k+\omega) f(s,t,\omega) \, .
\label{eq:I0}
\eeq

As mentioned above the precise kinematic bounds on the $\omega$ and $t$ integrals will naturally 
arise from the condition $g = B^2-A^2 \geq 0$. We easily obtain from \eq{AB}
\beq
g(\omega) = -a^2\omega^2 + b\,\omega +c \, ,
\label{eq:g(omega)}
\eeq
whose coefficients can be expressed as
\beqa
a &=& \frac{s-M^2}{p} \, , \nn  \\
b &=& -\frac{2t}{p^2} \l( E(s-M^2) - k(s+M^2) \r) ,\nonumber \\
c &=& -\frac{t}{p^2} \l[ t\l( (E+k)^2-s \r) + 4p^2k^2 -(s-M^2-2Ek)^2 \r] .
\label{abc}
\eeqa
Because the quadratic term in \eq{eq:g(omega)} is manifestly negative, $g(\omega)$ is 
positive in an interval $[\omega_{\rm min}, \omega_{\rm max}]$ where the discriminant $D = 4a^2c+b^2$ is positive. 
We have
\beqa
&& \ \ \ \omega_{\rm min}^{\rm max} = \frac{b \pm \sqrt{D}}{2a^2} \, , \\
&& D = -t \l( st+(s-M^2)^2 \r) \l( \frac{4k\sin\psi}p \r)^2 \, .
\label{D}
\eeqa
The condition $D \geq 0$ leads to the familiar range of the invariant momentum exchange in $2 \rightarrow 2$ 
processes with one massless and one massive collision partner, namely 
$t_{\rm min} \le t \le 0$ with 
\beq
t_{\rm min} = -\frac{(s-M^2)^2}{s}  \ \ .
\eeq

We now show that when $g(\omega) \geq 0$ (or equivalently 
$\omega_{\rm min} \leq \omega \leq \omega_{\rm max}$), the factor $\Theta(k+\omega)$ appearing
in \eq{eq:I0} is actually redundant. Recall that the condition $g(\omega) \geq 0$ arises from \eq{eq:phi-int}.
Thus the values of $\omega$ contributing to the l.h.s. of \eq{eq:phi-int}, \ie\ for which 
$\underline{K}'^2 = 2KQ+t =0$, must belong to the interval $[\omega_{\rm min}, \omega_{\rm max}]$. 
Such values thus satisfy
\beq
t + 2 k \omega = 2 {\bm k} {\bm q} \Rightarrow (t + 2 k \omega)^2 \leq 4 k^2 q^2 = 
4 k^2 (\omega^2 -t) \Rightarrow t(t+4k(k+\omega)) \leq 0 \ \ .
\eeq
Since $t \leq 0$ we find that $k+\omega \geq -t/(4k) \geq 0$. Thus the $\Theta(k+\omega)$ factor 
in \eq{eq:I0} can be dropped and we can now specify the precise bounds on $t$ and $\omega$:
\beq 
{\cal I}[f]=\frac1{16\pi^2 pE}\int_k \frac{n(k)}{2k}\int_{t_{\rm min}}^0 dt
\int_{\omega_{\rm min}}^{\omega_{\rm max}} \frac{d\omega}{\sqrt{g(\omega)}}\, \bar{n}(k+\omega) f(s,t,\omega) \, .
\label{eq:I}
\eeq

As a side remark, let us note that the change of variables \eq{eq:change_variables} 
maps the original integration area, $p' \in [0,\infty[$ and 
$\cos\theta \in [-1,+1]$, into a $(t,\omega)$ region enclosed by
\beq
\omega_\pm(t) = \frac{-t}{2M^2} \l( - E \pm p \sqrt{1-\frac{4M^2}t} \r) .
\label{eq:omega+-}
\eeq
>From this expression we check that the maximal energy transfer is as expected ${\rm Max}(\omega_+) = E-M$, 
occurring at $t = -2M(E-M)$ and corresponding to `full stopping'. Note also that our derivation of the bounds 
on $\omega$ in \eq{eq:I} implies that $\omega_{-}(t) \leq \omega_{\rm min} \leq \omega\leq \omega_{\rm max} \leq \omega_{+}(t)$.

\vskip 5mm
{\bf Approximation $\bar n \rightarrow 1$}
\vskip 5mm

The integrand in formula \eq{eq:I} contains the thermal distribution $\bar{n}(k+\omega)$, which usually 
prevents the calculation of the $\omega$-integral in terms of elementary functions.
We may, however, obtain useful approximations of the integral by replacing $\bar n = 1 \pm n \rightarrow 1$, \ie,
by neglecting thermal effects on the final states.

Under this assumption we can evaluate analytically the $\omega$-integral in \eq{eq:I} for the function 
$f(s,t,\omega) = \omega^\ell\, |{\cal M}|^2(s,t)$, for instance by using the formal identity
\beq
I_\omega^{(\ell)}=\int_{\omega_{\rm min}}^{\omega_{\rm max}} d\omega\,\frac{\omega^\ell}{\sqrt{g(\omega)}}=
{\rm Re} \int_{-\infty}^\infty d\omega\, \frac{\omega^\ell}{\sqrt{g(\omega)}} \, .
\eeq
Up to the prefactor $\Theta(D)$ involving the discriminant $D$ of the quadratic function $g(\omega)$, 
which reflects the 2-body kinematics and ensures a non-zero support of the integral as discussed above, 
we obtain for example
\beq
\begin{array}{c||c|c|c}
	\ell & 
		0 & 1 & 2
	\\ \hline
	I_\omega^{(\ell)} & 
		\;\D\frac\pi{a}\; & \;\D\frac\pi2\, \frac{b}{a^3}\; & 
		\;\D\frac\pi8\, \frac{D+2b^2}{a^5} \rule{0mm}{1.9em}
	\end{array} \medskip
\label{tableau}
\eeq
Thus, replacing $\bar n$ by unity in \eq{eq:I}, we find
\beq
	{\cal I}[f]
	\;\rightarrow\;
	{\cal I}^{(\ell)}[{\cal M}]
	=
	\frac1{16\pi^2 pE}
	\int_k \frac{n(k)}{2k}
	\int_{t_{\rm min}}^0 dt\, |{\cal M}|^2(s,t)\, I_\omega^{(\ell)} \, .
\eeq

For $\ell=1$, as of interest for the energy loss calculation, the approximation
$\bar n = 1 \pm n \rightarrow 1$ can be justified, as was done in Ref.~\cite{BT}, and as 
explained here below in Appendix C (after Eq.~\eq{loss1}). We will use this approximation, 
which allows to write, using \eq{tableau} for $\ell=1$ and \eq{abc},
\beq
{\cal I}^{(1)}[{\cal M}]=\int_k \frac{n(k)}{2k}\,\l( 1-\frac{s+M^2}{s-M^2}\frac{k}E \r)
\int_{t_{\rm min}}^0 dt\, (-t) 	\frac{|{\cal M}|^2(s,t)}{16\pi (s-M^2)^2} \, ,
\label{eq:Iapprox}
\eeq
which we recognize as an integral of the differential cross section weighted by the factor $t$. 

\section{Scattering off electrons: soft $q < q^{\star}$ contribution
\label{sec:appendixB}}

We start from Eq.~(40) of Ref.~\cite{BT}, which reads in the limit $v \to 1$:
\beq
-\l. \frac{dE_e}{dx} \r|_{\rm soft}=\frac{e^2}{8\pi} \int^{(q^\star)^2}_0 dq^2 \int_{-q}^{q} 
d\omega \, \omega \, \left[ \rho_L(\omega,q) + (1 -x^2) \rho_T(\omega,q) \right] \ \ ,
\label{soft1}
\eeq
where $x\equiv \omega/q$. For $v=1$, the $\omega$ integration range is the whole space-like region
$|\omega| \leq q$. We can shift to the time-like $|\omega| > q$ region by writing
\beq
\int_{-q}^{q} d\omega = \int_{-\infty}^{\infty} d\omega - \int_{|\omega| > q} d\omega  \, .
\label{shifttotimelike}
\eeq
In \eq{soft1} the resulting integral over the infinite $\omega$-range vanishes, which can easily be seen from the sum-rules\footnote{The sum rules can be derived from the spectral representations of the gluon propagators \cite{BI}
\beq
- \Delta_T(\omega, k) = \int dk_0 \frac{\rho_T(k_0, k)}{k_0-\omega} \, , \quad 
- \Delta_L(\omega, k) = - \frac{1}{k^2} + \int dk_0 \frac{\rho_L(k_0, k)}{k_0-\omega} \ , \nn
\eeq
by identifying in the $\omega \to \infty$ expansion of the left and right-hand sides the terms of appropriate order in $1/\omega$.}
\beqa
\int_{-\infty}^{\infty} d\omega \, \omega \, \rho_T(\omega, q) &=& 1 \, , \nn \\
\int_{-\infty}^{\infty} d\omega \, \omega^3 \, \rho_T(\omega, q) &=& q^2 + \frac{m_D^2}{3} \, , \nn \\
\int_{-\infty}^{\infty} d\omega \, \omega \, \rho_L(\omega, q) &=& \frac{m_D^2}{3 q^2}   \, ,
\label{sumrules} 
\eeqa
where $m_D^2 = e^2T^2/3$.

In the time-like region $|\omega| > q$, the spectral functions are given by the pole contributions ($s = L,T$)
\beq
\l. \rho_s(\omega,q)\r|_{|\omega| > q} = \epsilon(\omega) z_s(q) \delta(\omega^2 -\omega_s^2(q)) \, ,
\eeq
and \eq{soft1} becomes
\beq
-\l. \frac{dE_e}{dx} \r|_{\rm soft}=-\frac{e^2}{8\pi} \int^{(q^\star)^2}_0 dq^2 
\left[ z_L(q) + (1 -x_T^2) z_T(q) \right] \, , 
\label{soft2}
\eeq
where we denote $x_s \equiv \omega_s(q)/q$. 

By definition, the poles $\omega = \omega_s(q)$ of the propagators \eq{propagators} 
satisfy the implicit equations
\beq
q^2 = - \Pi_L(x_L) \, , \quad q^2 = \frac{\Pi_T(x_T)}{x_T^2-1} \, ,
\label{poles}
\eeq
with the longitudinal and transverse photon self-energies as specified in \eq{self}. 

The residues $z_s(q)$ are defined by 
\beq
\Delta_s(\omega, q) \simeq \frac{z_s(q)}{\omega^2 -\omega_s^2(q)} = \frac{z_s(q)}{q^2(x^2-x_s^2)} 
\ \ \ {\rm when} \ \ \ x^2 \simeq x_s^2(q) \equiv \omega_s^2(q)/q^2 \, .
\eeq
Expanding the denominators in \eq{propagators} around $x^2 \simeq x_s^2$ we get
\beqa
z_L(q)&=& 2x_L \frac{q^2}{\Pi'_L(x_L)} = -2x_L \frac{\Pi_L(x_L)}{\Pi'_L(x_L)} \, ,
 \nn \\
z_T(q)&=& \frac{1}{1-\frac{\Pi'_T(x_T)}{2q^2x_T}} = \frac{1}{1-\frac{x_T^2-1}{2 x_T}\frac{\Pi'_T(x_T)}{\Pi_T(x_T)}} \, .
\label{residues}
\eeqa
Thus $z_{L,T}(q)$ are explicit functions of $x_{L,T}$, which suggests to shift variables from $q^2$ to $x_{L,T}$ in \eq{soft2}. For the longitudinal and transverse contributions we find from \eq{poles} and \eq{residues}
\beqa
z_L(q) dq^2 &=& 2x_L \Pi_L(x_L) dx_L \, , \nn \\
(1-x_T^2) z_T(q) dq^2 &=& \frac{2 x_T}{x_T^2-1}\Pi_T(x_T) dx_T \, .
\eeqa
>From \eq{soft2} we then obtain
\beq
-\l. \frac{dE_e}{dx} \r|_{\rm soft}= \frac{e^2}{4\pi} \left\{ \int_{x_L(q^\star)}^{\infty} dx \, x \Pi_L(x)
+\int_{x_T(q^\star)}^{\infty} dx \, \frac{x \Pi_T(x)}{x^2-1} \right \} \, .
\label{soft3}
\eeq
Using now $q^\star \gg eT$ we have $x_{L,T}(q^\star) \simeq 1$, and since the first term of \eq{soft3} is integrable at $x=1$, we can safely replace $x_{L}(q^\star) \to x_{T}(q^\star)$ in this term. Using the relation $\Pi_T(x) = \Pi'_L(x) x (x^2-1)/2$ (obtained from \eq{self}), the two terms of \eq{soft3} combine into a full derivative, to give
\beq
-\l. \frac{dE_e}{dx} \r|_{\rm soft}= \frac{e^2}{4\pi} \left[ \frac{x^2}{2} \Pi_L(x)\right]_{x_T(q^\star)}^{\infty}
\mathop{\simeq}_{x_T \to 1}  \frac{e^2 m_D^2}{8 \pi} \left[-\frac{4}{3} +\frac{1}{2} \ln{\frac{2}{x_T(q^\star)-1}}\right] \, .
\eeq
Using finally $x_T(q^\star \gg eT) \simeq 1+m_D^2/(4 {q^\star}^2)$ \cite{BI} we arrive at
\beq
-\l. \frac{dE_e}{dx} \r|_{\rm soft}=  \frac{e^4 T^2}{24 \pi}\left[ \ln{\frac{q^\star}{e T}} +\frac{\ln 24}{2} -\frac{4}{3} \right]  \, .
\label{soft4}
\eeq

\section{Scattering off electrons: hard $q > q^{\star}$ contribution
\label{sec:appendixC}}

The hard contribution to $(-dE/dx)$ reads \cite{BT}
\beqa
\label{lossdef}
\left. -\frac{dE_e}{dx} \right|_{\rm hard} &=& \frac{1}{E} \int_{p'}\frac{1}{2E'}
\int_k \frac{n_F(k)}{2k} \int_{k'} \frac{{\bar n}_F(k')}{2k'} \nn \\
&&  \times (2\pi)^4 \delta^4(P+K-P'-K') \half \sum_{\rm spins} |{\cal M}|^2 \, \frac{\omega}{v}\, \Theta(q-q^\star) \, ,
\eeqa
where $\omega = E -E'$ is the energy transferred by the muon in the elastic scattering. 

The squared $t$-channel scattering amplitude (summed and averaged over spins) is given by
\beq
\label{Msquared}
\half \sum_{\rm spins} |{\cal M}|^2 = 16\,\frac{e^4}{t^2} 
\left[ (P K) (P' K') + (P K') (P' K) -M^2 K K'  \right] \ \ .
\eeq
In BT it is assumed that $k' \sim T$, which allows for the approximation
\beq
\label{MsquaredBT}
\half \sum_{\rm spins} |{\cal M}|^2 = 16\,\frac{e^4}{t^2} E E' 
\left[ 2(k- {\bm v} {\bm k})(k'- {\bm v} {\bm k}') + \frac{M^2 t}{2E^2} \right] \, ,
\eeq
where ${\bm v} = {\bm p}/E$ is the incoming muon velocity. 
However, \eq{Msquared} can easily be cast in a form similar to \eq{MsquaredBT} without any approximation.
Using $P' = P+K-K'$ and $K K' = - P Q = - t/2$ we obtain from \eq{Msquared} the exact 
expression
\beqa
\half \sum_{\rm spins} |{\cal M}|^2 &=& 16\,\frac{e^4}{t^2} 
\left[ 2 (P K) (P K') + (M^2 +t/2) t/2 \right] \nn \\
&=& 16\,\frac{e^4}{t^2} E^2 
\left[ 2(k- {\bm v} {\bm k})(k'- {\bm v} {\bm k}') + \frac{M^2 t}{2E^2} + \frac{t^2}{4 E^2} \right]  \ \ ,
\label{exactMsquared}
\eeqa
which differs from the BT approximation \eq{MsquaredBT} by the term $\propto t^2/E^2$ and the prefactor ($E^2$ instead of $E E'$). 

Recalling that $\omega \equiv k' - k$, the $\delta$-function for energy conservation in \eq{lossdef} can be expressed without approximation as
\beqa
\label{energycons}
\delta\left( E-E'-\omega \right)/(2E') &=& \delta\left( (E-\omega)^2 - E'^2 \right) = \delta\left( 2 {\bm p} {\bm q} +t -2 E\omega  \right) \nn \\
&=& \delta\left( \omega  -  {\bm v} {\bm q} - t/(2E) \right)/(2E) \, .
\eeqa
Using \eq{exactMsquared} and \eq{energycons}, and $3$-momentum conservation to perform the integral over ${\bm p}'$, the energy loss \eq{lossdef} becomes, again without any approximation
\beqa
\label{loss1}
\left. -\frac{dE_e}{dx} \right|_{\rm hard} &=& \frac{16\pi e^4}{v} 
\int_k \frac{n_F(k)}{2k} \int_{k'} \frac{{\bar n}_F(k')}{2k'}
\delta\left( \omega  -  {\bm v} {\bm q} - t/(2E) \right)  \nn \\
&&  \times   \frac{\omega}{t^2}\, \Theta(q-q^\star) 
\left[ 2(k- {\bm v} {\bm k})(k'- {\bm v} {\bm k}') + \frac{M^2 t}{2E^2} + \frac{t^2}{4 E^2} \right] \, .
\eeqa
This expression can be simplified as follows. First, as in the BT calculation, the term $\propto n_F(k')$ is neglected, \ie\ we replace ${\bar n}_F(k') \to 1$.
Indeed, for $k' \gg T$, $n_F(k')$ is exponentially suppressed. For $k' \sim T$, on the other hand, we have $\omega = k' - k \sim T$ and $q = |{\bm k}'-{\bm k}| \sim T$, thus $t/(2E) \sim T^2/E$ can be neglected in the $\delta$-function for energy conservation, and the term $\propto n_F(k')$ then vanishes by antisymmetry in $k \leftrightarrow k'$ (recall that $t = (K-K')^2$). Secondly, in the integrand of \eq{loss1} we insert 
\beq
1 = \int d^3 {\bm q} \, \delta^3({\bm q}+{\bm k}- {\bm k}') \int d\omega \, \delta(\omega +k-k') 
\eeq 
and perform the integral over ${\bm k}'$. This gives
\beqa
\label{loss2}
\left. -\frac{dE_e}{dx} \right|_{\rm hard} &=& \frac{4\pi e^4}{v} 
\int_k \frac{n_F(k)}{k} \int_q  \int d\omega \, \frac{\delta(\omega +k-|{\bm k}+{\bm q}|)}{|{\bm k}+{\bm q}|}  \,
\delta\left( \omega  - {\bm v} {\bm q} - t/(2E) \right)  \nn \\
&&  \times \frac{\omega}{t^2}\, \Theta(q-q^\star) 
\left[ 2(k- {\bm v} {\bm k})^2 + (k- {\bm v} {\bm k}) \frac{t}{E}+ \frac{M^2 t}{2E^2} + \frac{t^2}{4 E^2} \right] \, .
\eeqa
Eq.~\eq{loss2} differs from the BT expression only by the $t/(2E)$ term in the $\delta$-function for 
energy conservation and the $t^2/(4 E^2)$ term in the expression of the squared amplitude. 

As noted in BT, since $-dE/dx$ does not depend on the direction of ${\bm v}$, it is convenient to average \eq{loss2} over this direction, using
\beqa
\int \frac{d\Omega}{4 \pi} \delta (\tilde{\omega}  - {\bm v}{\bm q}) &=& 
\frac{\Theta(v^2q^2 -\tilde{\omega}^2)}{2vq} \, , \nn \\
\int \frac{d\Omega}{4 \pi} \delta (\tilde{\omega}  - {\bm v}{\bm q} ) \, v^i &=& 
\frac{\Theta(v^2q^2 -\tilde{\omega}^2)}{2vq} \frac{\tilde{\omega}}{q} \hat{q}^i \, , \\
\int \frac{d\Omega}{4 \pi} \delta (\tilde{\omega}  - {\bm v}{\bm q} ) \, v^i v^j &=& 
\frac{\Theta(v^2q^2 -\tilde{\omega}^2)}{2vq} 
\left[ \frac{v^2q^2 -\tilde{\omega}^2}{2q^2} \delta^{ij} + \frac{3\tilde{\omega}^2 - v^2q^2}{2q^2} \hat{q}^i \hat{q}^j \right] , \nn 
\eeqa
where we use the notation $\tilde{\omega} \equiv \omega  - t/(2E)$. Using then 
$\delta(\omega +k-|{\bm k}+{\bm q}|) = 2|{\bm k}+{\bm q}| \delta(t +2k \omega - 2 {\bm k}{\bm q})$ we find 
\beqa
\label{loss3}
\left. -\frac{dE_e}{dx} \right|_{\rm hard} &=& \frac{4\pi e^4}{v^2} 
\int_k \frac{n_F(k)}{k} \int_q \frac{1}{q}
\int_{\omega_{-}(q)}^{\omega_{+}(q)} d\omega \, \delta(t +2k \omega - 2 {\bm k}{\bm q})  \nn \\
&&  \times  \left[ 2 \left( k^2 -2 {\bm k}{\bm q} \frac{k \tilde{\omega}}{q^2} + \frac{v^2q^2 -\tilde{\omega}^2}{2q^2} k^2
+ \frac{3\tilde{\omega}^2 - v^2q^2}{2q^4} ({\bm k}{\bm q})^2 \right) \right. \nn \\
&& \left. + \left( k-{\bm k}{\bm q} \frac{\tilde{\omega}}{q^2} \right) \frac{t}{E}+ \frac{M^2 t}{2E^2} + \frac{t^2}{4 E^2} \right] 
\, \frac{\omega}{t^2} \Theta(q-q^\star) \, ,
\eeqa
where the factor $\Theta(v^2q^2 -\tilde{\omega}^2)$ yields the bounds on $\omega$, 
\beq
\omega_{\pm}(q) \equiv E - \sqrt{E^2 +q^2 \mp 2Evq} \, .
\label{ompm}
\eeq
>From \eq{loss3} we proceed as follows. We replace $2 {\bm k}{\bm q} \rightarrow t +2k \omega$ in the integrand, 
perform the integral over the angle between ${\bm k}$ and ${\bm q}$ using
\beq
\int_{-1}^{1} d \cos{\theta} \, \delta(t +2k \omega - 2 k q \cos{\theta}) = \Theta(|q-k|\leq |\omega +k| \leq q+k) /(2kq) \ \ ,
\eeq
and then re-express $\tilde{\omega} \to \omega  - t/(2E)$. Ordering in powers of $1/E$, we obtain
\beqa
\left. -\frac{dE_e}{dx} \right|_{\rm hard} &=& \frac{e^4}{4 \pi^3 v^2} 
\int_0^{\infty} dk \, n_F(k) \int_{q^\star}^{\infty} dq \int_{\omega_{-}(q)}^{\omega_{+}(q)} d\omega \, 
\Theta(|q-k|\leq |\omega +k| \leq q+k) \nn \\
&& \times \frac{\omega}{q^2} \, \left\{ \frac{3\omega^2}{4q^2} - \frac{v^2}{4} + \frac{1-v^2}{2} \frac{q^2}{t} + 3 \frac{k(k+\omega)}{q^2}
+ (1-v^2)\frac{k(k+\omega)}{t} \right. \nn \\
&&  \left. \hskip -2cm  -\frac{\omega \left[ 12 k(k+\omega) + 3 \omega^2 -q^2 \right]}{4 q^2 E} 
+ \frac{4 k(k+\omega)(3 \omega^2 -q^2) + 3 \omega^4 -2 \omega^2 q^2 - q^4}{16 q^2 E^2}
+\frac{q^2}{4 E^2} \right\} . \nn \\
\label{loss4}
\eeqa
We mention here that all $1/E$ and $1/E^2$ terms stem from our `correction' $t/(2E)$ in the $\delta$-function of \eq{loss2}, except the last term $\sim q^2/(4 E^2)$, which comes from the exact expression \eq{exactMsquared}
of the squared amplitude. We have separated these terms to underline (see \eq{loss7} and \eq{correctedloss}) that the error made in \cite{BT} is due to an incorrect approximation both in the $\delta$-function for energy conservation and in the expression of the squared amplitude. 

Using $-q \leq \omega_{-}(q) \leq \omega_{+}(q) \leq q$ we can show that the phase space constraints in \eq{loss4} can be written as 
\beqa
&& \Theta(q-q^\star) \, \Theta(|q-k|\leq |\omega +k| \leq q+k) \, \Theta(\omega_{-} \leq \omega \leq \omega_{+}) = \nn \\
&& \hskip 1cm \Theta({\bar k}-q^\star) \, \Theta(q^\star \leq q \leq {\bar k}) \, \Theta(\omega_{-} \leq \omega \leq \omega_{+}) \nn \\
&& \hskip 1cm + \Theta({\rm Max}({\bar k},q^\star) \leq q \leq q_{\rm max}) \, \Theta(q-2k \leq \omega \leq \omega_{+}) \, ,
\eeqa
where ${\bar k}$ and $q_{\rm max}$ are the values of $q$ at which $q-2k =\omega_{-}(q)$ and  $q-2k =\omega_{+}(q)$,
\beq 
\label{qmax}
{\bar k} = \frac{2k(E+k)}{E(1+v)+2k} \, , \qquad  q_{\rm max} = \frac{2k(E+k)}{E(1-v)+2k} \, .
\eeq
At this point we use $k \sim T \ll E$ to approximate ${\bar k} \simeq 2k /(1+v)$, and $q^\star \ll T$ so that effectively ${\bar k} > q^\star$. Also, in the part of the integral $\propto \Theta(q^\star \leq q \leq {\bar k})$ 
we can approximate $\omega_{\pm}(q) \simeq \pm vq$. We can thus replace in \eq{loss4}
\beq
\int_{q^\star}^{\infty} dq \int_{\omega_{-}(q)}^{\omega_{+}(q)} d\omega \, \Theta( \ )
\to \left[ \int_{q^\star}^{2k/(1+v)} dq \int_{-v q}^{v q} d\omega 
+ \int_{2k/(1+v)}^{q_{\rm max}} dq \int_{q-2k}^{\omega_{+}(q)} d\omega \right] \, .
\label{shiftinloss4}
\eeq
In the term corresponding to $q^\star \leq q \leq 2k/(1+v)$, $q$ and $\omega$ are constrained to be of order $T$. It is then 
easy to see that in the curly bracket of \eq{loss4}, the first line will contribute as $\sim e^4T^2$ to the energy loss, whereas the 
second line can be neglected, since these terms are suppressed by $\morder{T/E}$ and $\morder{T^2/E^2}$, respectively. 
For the contribution from $2k/(1+v) \leq q \leq q_{\rm max}$ in \eq{shiftinloss4}, 
the $\omega$-range brings a factor $\sim k \sim T$, as can be seen from the identity
\beq
\omega_{+}(q) - (q-2k) = \frac{4k(E+k)}{2(E+k)-q-\omega_{+}(q)} \, \left(1-\frac{q}{q_{\rm max}} \right) \ \ .
\eeq
When $q \sim q_{\rm max} \gg T$, we have there $\omega \sim q \sim q_{\rm max}$, and the second line of the curly bracket of \eq{loss4}
contributes to $\morder{q_{\rm max}/E}$ and $\morder{q_{\rm max}^2/E^2}$ (the terms $\propto k$ are suppressed by at least $\morder{T/E}$ and 
can be dropped). This contribution is thus important when 
$q_{\rm max} \sim E$. From \eq{qmax} this happens when $E(1-v) \lesssim k \sim T$, \ie\ when $E \gsim M^2/T$. 

Using \eq{shiftinloss4} the expression \eq{loss4} can be written as
\beqa
&& \left. -\frac{dE_e}{dx} \right|_{\rm hard} = \frac{e^4}{4 \pi^3 v^2}
\int_0^{\infty} dk \, n_F(k) \left[ \int_{q^\star}^{2k/(1+v)} \frac{dq}{q^2} \int_{-v q}^{v q} d\omega\,\omega
+ \int_{2k/(1+v)}^{q_{\rm max}} \frac{dq}{q^2} \int_{q-2k}^{\omega_{+}(q)} d\omega \,\omega \right] \nn \\
&& \hskip 3cm \times \left\{ \frac{3\omega^2}{4q^2} - \frac{v^2}{4} + \frac{1-v^2}{2} \frac{q^2}{Q^2} + 3 \frac{k(k+\omega)}{q^2} + (1-v^2)\frac{k(k+\omega)}{Q^2} \right\} \nn \\
&& + \frac{e^4}{4 \pi^3 v^2}\!
\int_0^{\infty}\!\! dk \, n_F(k)\!
\int_{2k/(1+v)}^{q_{\rm max}} \frac{dq}{q^2} \int_{q-2k}^{\omega_{+}(q)}\!\! d\omega\,\omega
\left\{\frac{\omega(q^2 - 3 \omega^2)}{4 q^2 E} + \frac{3 \omega^4 -2 \omega^2 q^2 - q^4 }{16 q^2 E^2} +\frac{q^2}{4 E^2} \right\}. \nn \\
\label{loss6}
\eeqa
This expression differs from the BT pendant by the upper bound $\omega_{+}(q)$ (instead of $v q$) in the first line, and most importantly by the presence of the third line, containing terms formally $\propto 1/E$ and $\propto 1/E^2$. In order to compare further to the BT calculation, 
let us consider as in \cite{BT} the two limiting cases $E \ll M^2/T$ and $E \gg M^2/T$, where the expression 
\eq{qmax} of $q_{\rm max}$ can be approximated as 
\beq
\label{qmaxappr}
q_{\rm max} \mathop{\simeq}_{E \ll \frac{M^2}{T}} \frac{2k}{1-v} \sim \frac{E^2}{M^2/T} \ll E \quad \mbox{vs.} \quad
q_{\rm max} \mathop{\simeq}_{E \gg \frac{M^2}{T}} E \, .
\eeq
In the domain $E \ll M^2/T$, we have $q \leq q_{\rm max} \ll E$, thus $\omega_{+}(q)\simeq vq$ from \eq{ompm}. For $E \gg M^2/T$, we can approximate
\beq
\label{omappr}
\omega_{+}(q) =  E - \sqrt{E^2 +q^2 - 2Evq} = E-\sqrt{(E-q)^2 +\frac{2M^2}{1+v} \frac{q}{E}} \simeq q \, ,
\eeq
where we assumed $E-q \gg M$. This is justified since the contribution 
from $E-M \leq q \leq E$ to the energy loss \eq{loss6} is of order $e^4 T^2 M /E$ and thus suppressed 
compared to the dominant contribution $\sim e^4T^2$ we are looking for. Thus either when $E \ll M^2/T$, 
or when $E \gg M^2/T$ (corresponding to the ultrarelativistic $v \to 1$ limit), the approximation $\omega_{+}(q)\simeq vq$ in \eq{loss6} is valid\footnote{This approximation would be incorrect in the 
intermediate regime $E \sim M^2/T$, where the exact expression \eq{ompm} of $\omega_{+}(q)$ should be used.}, 
and the only difference between our result and the 
BT calculation is the additional term written in the third line of \eq{loss6}. As discussed previously, this 
term contributes to the energy loss as $\sim e^4 T^2$ only when $E \gsim M^2/T$. We conclude that 
the domain $E \gg M^2/T$ (\ie\ $v \to 1$) is treated incorrectly in \cite{BT}. 
Focusing now on this limit, we obtain from \eq{loss6}
\beqa
\left. -\frac{dE_e}{dx} \right|_{\rm hard}^{v\to 1} = \left. -\frac{dE_e}{dx} \right|_{\rm hard, BT}^{v\to 1} + \frac{e^4}{4 \pi^3} \int_0^{\infty} dk \, n_F(k) \int_{k}^{E} \frac{dq}{q^2} \int_{q-2k}^{q} d\omega\,\omega 
\left\{-\frac{q}{2 E} + \frac{q^2}{4 E^2}  \right\} , \nn \\
\label{loss7}
\eeqa
where we replaced $\omega \to q$ in the bracket of the $\omega$ integral, the terms $\propto (q-\omega)$ yielding
negligible contributions when $E \to \infty$. As mentioned after \eq{loss4}, the corrections to the BT result arise from using the exact squared amplitude \eq{exactMsquared} (term $\sim q^2/(4 E^2)$ in \eq{loss7}) and the 
$\delta$-function for exact energy conservation (additional term $\sim -q/(2 E)$). 
The remaining integrals in \eq{loss7} are trivial and we get to leading order
in $1/E$
\beq
\left. -\frac{dE_e}{dx} \right|_{\rm hard}^{v\to 1} = \left. -\frac{dE_e}{dx} \right|_{\rm hard, BT}^{v\to 1} 
+\frac{e^4 T^2}{48 \pi} \left( - \frac{3}{4}\right) .
\label{correctedloss}
\eeq
Our new term in \eq{correctedloss} arises from a kinematical domain
where the momentum exchange $q$ is `very hard', $q \sim q_{\rm max} \simeq E$. 
This domain already contributed to the BT result (written in \eq{BThard} below). Indeed, the logarithmic 
term $\propto \ln{E/T}$ arises from an integral $\sim \int^E_T dq/q$, where, for instance, the interval $E/2 \leq q \leq E$ contributes as $\ln{2}$. 
The very hard region was however not consistently treated in \cite{BT}, due to the ad hoc use of the approximation $q \ll E$. 

Given that \cite{BT} 
\beq
\left. -\frac{dE_e}{dx} \right|_{\rm hard, BT}^{v \to 1} = \frac{e^4T^2}{48\pi} 
\left[ \ln{\frac{2TE}{(q^\star)^2} + \frac{8}{3} - \gamma +\frac{\zeta'(2)}{\zeta(2)}} \right] \, ,
\label{BThard}
\eeq
our result \eq{correctedloss} reads
\beq
\left. -\frac{dE_e}{dx} \right|_{\rm hard}^{v\to 1} = \frac{e^4 T^2}{48 \pi} 
\left[ \ln{\frac{2TE}{(q^\star)^2}}  + \frac{8}{3} - \gamma + \frac{\zeta'(2)}{\zeta(2)}  - \frac{3}{4} \right]  \, .
\label{totalhard}
\eeq

\providecommand{\href}[2]{#2}\begingroup\raggedright

\endgroup


\begin{thebibliography}{10}

\bibitem{bj}
  J.~D.~Bjorken,
  %``Energy Loss Of Energetic Partons In Quark - Gluon Plasma: Possible
  %Extinction Of High P(T) Jets In Hadron - Hadron Collisions,''
Fermilab preprint PUB-82/59-THY (1982).

\bibitem{phenix} 
  K.~Adcox {\it et al.} [PHENIX Collaboration],
  %``Suppression of hadrons with large transverse momentum in central  Au + Au
  %collisions at s**(1/2)(N N) = 130-GeV,''
  Phys.\ Rev.\ Lett.\  {\bf 88} (2002) 022301 [arXiv:nucl-ex/0109003]; 
%  S.~S.~Adler {\it et al.}  [PHENIX Collaboration],
%  %``Absence of suppression in particle production at large transverse  momentum
%  %in s(NN)**(1/2) = 200-GeV d + Au collisions,''
%  Phys.\ Rev.\ Lett.\  {\bf 91} (2003) 072303 [arXiv:nucl-ex/0306021].
  S.~S.~Adler {\it et al.}  [PHENIX Collaboration],
  %``Suppressed pi0 production at large transverse momentum in central Au +  Au
  %collisions at s(NN)**(1/2) = 200-GeV,''
  Phys.\ Rev.\ Lett.\  {\bf 91} (2003) 072301 [arXiv:nucl-ex/0304022].

\bibitem{star}
  C.~Adler {\it et al.}  [STAR Collaboration],
  %``Centrality dependence of high p(T) hadron suppression in Au + Au collisions
  %at s(NN)**(1/2) = 130-GeV,''
  Phys.\ Rev.\ Lett.\  {\bf 89} (2002) 202301
  [arXiv:nucl-ex/0206011].
  %%CITATION = NUCL-EX 0206011;%%

%\cite{Adler:2005xv}
\bibitem{Adler:2005xv}
  S.~S.~Adler {\it et al.}  [PHENIX Collaboration],
  %``Nuclear modification of electron spectra and implications for heavy quark
  %energy loss in Au + Au collisions at s(NN)**(1/2) = 200-GeV,''
  Phys.\ Rev.\ Lett.\  {\bf 96} (2006) 032301
  [arXiv:nucl-ex/0510047].
  %%CITATION = NUCL-EX 0510047;%%

%\cite{Bielcik:2005wu}
\bibitem{Bielcik:2005wu}
  J.~Bielcik  [STAR Collaboration],
  %``Centrality dependence of heavy flavor production from single electron
  %measurement in s(NN)**(1/2) = 200-GeV Au + Au collisions,''
  Nucl.\ Phys.\  A {\bf 774}, 697 (2006)
  [arXiv:nucl-ex/0511005].
  %%CITATION = NUPHA,A774,697;%%

%\cite{Armesto:2005mz}
\bibitem{Armesto:2005mz}
  N.~Armesto, M.~Cacciari, A.~Dainese, C.~A.~Salgado and U.~A.~Wiedemann,
  %``How sensitive are high-p(T) electron spectra at RHIC to heavy quark  energy
  %loss?,''
  Phys.\ Lett.\  B {\bf 637}, 362 (2006)
  [arXiv:hep-ph/0511257].
  %%CITATION = PHLTA,B637,362;%%

%\cite{Wicks:2005gt}
\bibitem{Wicks:2005gt}
  S.~Wicks, W.~Horowitz, M.~Djordjevic and M.~Gyulassy,
  %``Elastic, Inelastic, and Path Length Fluctuations in Jet Tomography,''
  Nucl.\ Phys.\  A {\bf 784}, 426 (2007)
  [arXiv:nucl-th/0512076].
  %%CITATION = NUPHA,A784,426;%%

\bibitem{TG}
  M.~H.~Thoma and M.~Gyulassy,
  %``Quark Damping And Energy Loss In The High Temperature QCD,''
  Nucl.\ Phys.\ B {\bf 351} (1991) 491.
  %%CITATION = NUPHA,B351,491;%%

\bibitem{BTqcd}
  E.~Braaten and M.~H.~Thoma,
  %``Energy loss of a heavy fermion in a hot plasma,''
  Phys.\ Rev.\ D {\bf 44} (1991) 2625.



\bibitem{PGG}
  S.~Peign\'e, P.~B.~Gossiaux and T.~Gousset,
  %``Retardation effect for collisional energy loss of hard partons produced in
  %a QGP,''
  J. High Energy Phys. JHEP04(2006)011 [arXiv:hep-ph/0509185].
  %%CITATION = HEP-PH 0509185; 	

\bibitem{Djordjevic:2006tw}
  M.~Djordjevic,
  %``Collisional energy loss in a finite size QCD matter,''
  Phys.\ Rev.\ C {\bf 74} (2006) 064907
  [arXiv:nucl-th/0603066].
  %%CITATION = NUCL-TH 0603066;%%%

%\cite{Wang:2006qr}
\bibitem{Wang:2006qr}
  X.~N.~Wang,
  %``Interference effect in elastic parton energy loss in a finite medium,''
  Phys.\ Lett.\  B {\bf 650}, 213 (2007)
  [arXiv:nucl-th/0604040].
  %%CITATION = PHLTA,B650,213;%%

%\cite{Gossiaux:2006yr}
\bibitem{Gossiaux:2006yr}
  P.~B.~Gossiaux, S.~Peign\'e, C.~Brandt and J.~Aichelin,
  %``Energy Loss of a Heavy Quark Produced in a Finite Size Medium,''
  JHEP {\bf 0704}, 012 (2007)
  [arXiv:hep-ph/0608061].
  %%CITATION = JHEPA,0704,012;%%

%\cite{Gossiaux:2007gd}
\bibitem{Gossiaux:2007gd}
  P.~B.~Gossiaux, J.~Aichelin, C.~Brandt, T.~Gousset and S.~Peign\'e,
  %``Energy loss of a heavy quark produced in a finite-size quark-gluon
  %plasma,''
  J.\ Phys.\ G {\bf 34}, S817 (2007)
  [arXiv:hep-ph/0703095].
  %%CITATION = JPHGB,G34,S817;%%

\bibitem{BT}
  E.~Braaten and M.~H.~Thoma,
  %``Energy loss of a heavy fermion in a hot plasma,''
  Phys.\ Rev.\ D {\bf 44} (1991) 1298.

\bibitem{Zakharov}
  B.~G.~Zakharov,
  %``Parton energy loss in an expanding quark-gluon plasma: Radiative vs
  %collisional,''
  arXiv:0708.0816 [hep-ph].
  %%CITATION = ARXIV:0708.0816;%%

\bibitem{Qin}
  G.~Y.~Qin, J.~Ruppert, C.~Gale, S.~Jeon, G.~D.~Moore and M.~G.~Mustafa,
  %``Radiative and Collisional Jet Energy Loss in the Quark-Gluon Plasma at
  %RHIC,''
  arXiv:0710.0605 [hep-ph].
  %%CITATION = ARXIV:0710.0605;%%

%\cite{DuttMazumder:2004xk}
\bibitem{alam}
  A.~K.~Dutt-Mazumder, J.~e.~Alam, P.~Roy and B.~Sinha,
  %``Stopping power of hot QCD plasma,''
  Phys.\ Rev.\  D {\bf 71} (2005) 094016
  [arXiv:hep-ph/0411015].
  %%CITATION = PHRVA,D71,094016;%%

\bibitem{Weldon:1983jn}
  H.~A.~Weldon,
  %``Simple Rules For Discontinuities In Finite Temperature Field Theory,''
  Phys.\ Rev.\  D {\bf 28}, 2007 (1983).

%\cite{Braaten:1991dd}
\bibitem{Braaten:1991dd}
  E.~Braaten and T.~C.~Yuan,
  %``Calculation of screening in a hot plasma,''
  Phys.\ Rev.\ Lett.\  {\bf 66} (1991) 2183.
  %%CITATION = PRLTA,66,2183;%%

\bibitem{pisarski}
  R.~D.~Pisarski,
  %``Renormalized Gauge Propagator In Hot Gauge Theories,''
  Physica A {\bf 158} (1989) 146.
  %%CITATION = PHYSA,A158,146;%%%
  E.~Braaten and R.~D.~Pisarski,
  %``Resummation And Gauge Invariance Of The Gluon Damping Rate In Hot QCD,''
  Phys.\ Rev.\ Lett.\  {\bf 64} (1990) 1338; Nucl.\ Phys.\ B {\bf 337} (1990) 569; Nucl.\ Phys.\ B {\bf 339} (1990) 310.
  
\bibitem{BI}
  J.~P.~Blaizot and E.~Iancu,
  %``The quark-gluon plasma: Collective dynamics and hard thermal loops,''
  Phys.\ Rept.\  {\bf 359} (2002) 355
  [arXiv:hep-ph/0101103].
  %%CITATION = HEP-PH 0101103;%%

\bibitem{Peshier:1998dy}
  A.~Peshier, K.~Schertler and M.~H.~Thoma,
  %``One-loop self energies at finite temperature,''
  Annals Phys.\  {\bf 266}, 162 (1998) [arXiv:hep-ph/9708434].

%\cite{Peskin:1995ev}
\bibitem{PS}
  M.~E.~Peskin and D.~V.~Schroeder, `An Introduction To Quantum Field Theory', 
%\href{http://www.slac.stanford.edu/spires/find/hep/www?irn=3485960}{SPIRES entry}
{\it  Reading, USA: Addison-Wesley (1995)}.


\bibitem{Peshier:2006hi}
  A.~Peshier,
  %``The QCD collisional energy loss revised,''
  Phys.\ Rev.\ Lett.\  {\bf 97}, 212301 (2006)
  [arXiv:hep-ph/0605294].

\end{thebibliography}
\end{document}